\documentclass[]{emulateapj}
\usepackage{epsfig}
\usepackage{amssymb,amsmath}
\usepackage{natbib}
\usepackage{multirow}
\usepackage{color}			      % using color package        - for emulateapj

\def\bb{$B_{\rm 435}\,$} % ACS B filter F435W 
\def\vv{$V_{\rm 606}\,$} % ACS V filter F606W
\def\ii{$i_{\rm 775}\,$} % ACS i filter F775W
\def\zz{$z_{\rm 850}\,$} % ACS z filter F850W
\def\i814{$I_{\rm 814}\,$} % ACS I filter F814W
\def\y098{$Y_{\rm 098}$} %WFC3/IR Y098 filter
\def\jj{$J_{\rm 125}\,$} %WFC3/IR J filter  
\def\yy{$Y_{\rm 105}\,$} %WFC3/IR Y filter  
\def\hh{$H_{\rm 160}\,$} %WFC3/IR H filter  
\newcommand{\myemail}{benedetta.vulcani@unimelb.edu.au}

\shorttitle{Interlopers in high redshift samples}
\shortauthors{Vulcani B. et al.}
\begin{document}

\title{Characterization and modeling of contamination for Lyman break
  galaxy samples at high redshift} 
  
 \author{Benedetta
  Vulcani\altaffilmark{1 \dag}, Michele Trenti\altaffilmark{1},
  Valentina Calvi\altaffilmark{2}, Rychard Bouwens\altaffilmark{3,4},
  Pascal Oesch\altaffilmark{5}, Massimo Stiavelli\altaffilmark{2}, Marijn Franx\altaffilmark{3}}
\altaffiltext{\dag}{\myemail} \affil{\altaffilmark{1}School of
  Physics, Tin Alley, University of Melbourne VIC 3010, Australia }
\affil{\altaffilmark{2} Space Telescope Science Institute, Baltimore,
  MD, 21218, USA}
\affil{\altaffilmark{3}Leiden Observatory, Leiden University, NL-2300 RA Leiden, Netherlands}
\affil{\altaffilmark{4}UCO/Lick Observatory, University of California, Santa Cruz, CA 95064, USA}
\affil{\altaffilmark{5}Yale Center for Astronomy and Astrophysics,
  Yale University, New Haven, CT 06511, USA}
\begin{abstract}
  The selection of high redshift sources from broad-band photometry
  using the Lyman-break galaxy (LBG) technique is a well established methodology,
  but the characterization of its contamination for the faintest
  sources is still incomplete. We use the optical and near-IR
  data from four (ultra)deep Hubble Space Telescope legacy fields to
  investigate the contamination fraction of LBG samples at
  $z\sim5-8$ selected using a colour-colour method. Our approach is based on characterizing the number count
  distribution of interloper sources, that is galaxies with colors
  similar to those of LBGs, but showing detection at wavelengths
  shorter than the spectral break.  Without sufficient sensitivity at
  bluer wavelengths, a subset of interlopers may not be properly
  classified, and  contaminate the LBG selection.  
  The surface density of interlopers in the sky gets steeper  
  with increasing redshift of LBG selections. Since the
  intrinsic number of dropouts decreases significantly with increasing
  redshift, this implies increasing contamination from misclassified
  interlopers with increasing redshift, primarily by intermediate
  redshift sources with unremarkable properties (intermediate ages,
  lack of ongoing star formation and low/moderate dust content). Using
  Monte Carlo simulations, we estimate that the CANDELS deep data have
  contamination induced by photometric scatter increasing from $\sim2\%$
  at $z\sim 5$ to $\sim 6\%$ at $z\sim 8$ for a typical dropout color
  $\geq 1$ mag, with contamination naturally decreasing for a more stringent
  dropout selection. Contaminants are expected to be located
  preferentially near the detection limit of surveys, 
  ranging from 0.1 to 0.4 contaminants per arcmin$^2$ at \jj=30,
  depending on the field considered.
 % reaching
 % fractions as high as $\sim 20-30\%$ for $m_{AB}\sim 29$ sources at
 % $z\sim 6-8$ in ultradeep surveys like the XDF.  
  This analysis suggests that the impact of
  contamination in future studies of $z > 10$ galaxies needs to be
  carefully considered.
 \end{abstract}

\keywords{cosmology: observations --- galaxies: evolution --- galaxies: high-redshift ---
  galaxies: photometry}

\section{Introduction}
\label{sec:introduction}

The Lyman-break technique, first proposed by \citet{steidel96},
transformed the identification of reliable samples of galaxy
candidates at high-redshift from broad-band imaging, and it is now
routinely used to study galaxy formation and evolution as early as
$500$ Myr after the Big Bang, at redshift $z\sim 10$ (e.g., see
\citealt{coe2015, bouwens2015, mcleod2016, oesch2016}). While one could consider selecting high redshift samples based on the best-fit 
photometric redshift or redshift likelihood contours 
\citep[e.g.,][]{mclure2010, finkelstein2012, bradley2014}, Lyman break selection 
procedures utilising cuts in colour space can be simpler to apply and offer a slight advantage in terms of operational transparency. 
This makes such a selection procedure easier to reproduce by both theorists and observers, 
as follow-up studies by \cite{shimizu2014, lorenzoni2013, schenker2013} 
 show.

The idea of the method rests on the
identification of the strong spectral break introduced by neutral
hydrogen atoms along the line of sight at wavelengths shorter than
Lyman-$\alpha$ (1216 \AA\/ rest-frame).\footnote{Note that there is a
  further suppression of the flux in the region across the 912 \AA\/
  rest-frame Lyman-continuum discontinuity, but in practice for
  galaxies at $z\gtrsim 6$ the non-detection starts at
  $\lambda \leq 1216$ \AA\/ rest-frame.} Thus, to identify probable
sources at high redshift with high confidence, the Lyman-break
selection typically resorts to three crucial ingredients: (1) color
information from two adjacent passbands to locate the wavelength
location and measure the amplitude of the break, (2) color information
red-ward of the break to characterize the intrinsic color of the
source, and (3) evidence that sources have no flux blueward of the
break.

Many studies have used different color selections, also depending on the availability  
of the photometric bands \citep[e.g.][to cite a few]{giavalisco2004, bouwens2007,
castellano2012, bradley2012, oesch2012, oesch2014, bouwens2011,bouwens2015}, 
showing how different choices can still lead to comparable results and assessing the strength of the method.

The  Lyman-break technique has been applied very successfully to build large
samples of galaxies, especially from Hubble Space Telescope (\emph{HST})
imaging (e.g. more than 10000 sources identified at
$3.5 \lesssim z \lesssim 11$ from \emph{HST} legacy fields to date; see
\citealt{bouwens2015}). Also, substantial spectroscopic follow-up work
has shown that samples are generally reliable, and that contamination from
sources with similar colors but different redshift is generally under
control \citep{steidel1999, bunker2003, malhotra2005, dowhygelund2007,
  popesso2009, vanzella2009, stark2010}. Nonetheless, photometrically
defined samples are intrinsically affected by contamination. While
this possibility is universally acknowledged in the literature and
specific studies estimate the contamination rate of the samples
presented (e.g. \citealt{su2011,pirzkal2013,bouwens2015}),
surprisingly few studies have been devoted to a detailed
characterization of the contamination rate and of its dependence on
survey parameters and redshift of the galaxy population. Potential
classes of contaminants that have been identified include stellar sources, low
redshift galaxies with prominent 4000~\AA/Balmer breaks and dust,
extreme emission line galaxies, time-variable sources such as Supernovae,
with the first two classes of objects representing the major risks
\citep{stanway2008, atek2011,bowler2012}.

Dwarf stars have similar colors to high-redshift galaxies because of
their low surface temperatures, and can thus enter dropout samples,
especially at $z\geq 7$ in data that lack sufficient angular resolution to
discriminate point sources from extended light profiles
\citep{stanway2003,bouwens2006,ouchi2009b,tilvi2013,wilkins2014}. 
At these redshifts, very low temperatures  stars (sub-types M, L, T and Y) result in sources that are intrinsically 
faint, and spectra in which the continuum is interrupted may
show large breaks across narrow-wavelength ranges, or in which the
flux peaks in narrow regions.
While deep medium-band observations are efficient in identifying these
stellar contaminants in seeing-limited ground-based observations \citep{wilkins2014}, \emph{HST} imaging is generally effective in
identifying stellar objects that are detected at signal-to-noise $S/N\gtrsim 10$
\citep{finkelstein2010,bouwens2011z78}. In addition, we note that at $z>9$,  the contamination
from stars is negligible, since there are essentially no observed stars with spectral energy distributions (SEDs) that peak
at $>1.4\mu m$ and are undetectable in the optical for typical HST
surveys \citep[e.g.][]{oesch2014}.

The main source of contamination for space-based surveys is thus that
of low/intermediate redshift galaxies that have a deep break around
4000 \AA{} rest-frame. The nature of these contaminants has not been
investigated in detail, but likely they are low-mass, moderate-age,
Balmer break galaxies at $z\sim 1-3$ \citep{wilkins2010, hayes2012}, possibly
with strong emission lines that contribute, or even dominate, the flux
redward of the spectral break \citep{vanderwel2011, atek2011}. To
effectively discriminate between the high-$z$ Lyman-break and the 4000~\AA{}/Balmer
break, \citet{stanway2008} recommend using a set of non-overlapping,
but adjacent, filters, so that a clear color cut can be imposed on the
selection. Another key requirement to build a clean sample is the
availability of very deep observations blueward of the spectral break,
to distinguish between a true non-detection for an high-$z$ object, and
a faint continuum for an interloper \citep{bouwens2015}. 

The goal of this paper is to focus on this class of intermediate
redshift interlopers, and to empirically quantify their impact on
high-$z$ Lyman-break galaxy (LBG) samples selected via a colour-cut method and characterize how their incidence
varies with depth and adopted selection cut. 
 For this, we resort to the optical and
near-infrared imaging on the GOODS South deep, GOODS North wide fields observed by the
CANDLES program \citep{grogin2011} and the XDF \citep{illingworth2013} and HUDF09-2 \citep{bouwens2011_hudf09}
fields. These datasets provide us high-quality
multi-wavelength observations over different 
areas of the sky (from $\sim 4.7$  arcmin$^2$ to $\sim 64.5$ arcmin$^2$). Specifically, we focus on LBG samples from
$z\sim5$ to $z\sim8$, and investigate the population of galaxies that
satisfy the color-color requirements to be included in the LBG
selection based on imaging at wavelengths starting from the spectral
break, but show a clear detection in bluer filters. We define this
class of objects as interlopers, and characterize (1) their surface
density in the sky depending on luminosity and on the redshift of the
dropout selection; (2) the likelihood that fainter counterparts of the
known population of interlopers enter a LBG sample because of lack of
sufficiently deep imaging in the blue.  We define this population as contaminants.

The results of our analysis, based on some of the deepest Hubble
observations available, have multiple applications. In particular,
they find applications to the estimation of the contamination rate of
other surveys, which may lack the multi-wavelength, multi-observatory
coverage, such as random pointings
and/or parallel observations (e.g. see
\citealt{trenti2011,trenti2012,bradley2012,schmidt2014,calvi2016}). Another
important application includes planning and optimization of future
observations (e.g., see \citealt{mason2015} for JWST and WFIRST
surveys at high-$z$).

This paper is organized as
follows: in Section \ref{Sec:dataset} we introduce our dataset and %, and in
%Section \ref{Sec:sample} we 
construct the samples of dropouts and
interlopers. In Section \ref{Sec:characterization_contaminants} we
analyze and discuss the properties of the contaminants and the
expected impact on LBG samples. In Section \ref{Sec:lit} we discuss
how results depend on the selection criteria. We summarize and conclude in Section
\ref{Sec:conclusions}.  Throughout the paper, we assume
$\Omega_0 = 0.3$, $\Omega_\Lambda = 0.7$, and
$H_0 = 70\, km\, s^{-1}\, Mpc^{-1}$. All magnitudes are in the AB
system \citep{oke1983}.

\section{Dataset and sample selection}
\label{Sec:dataset}

We base our analysis on four different samples, in order to test how results change
with the field  used for selection. We use the  CANDELS/GOODS South deep  (GSd) and 
CANDELS/GOODS North wide (GNw) imaging
\citep{grogin2011}, the entire XDF data set \citep{illingworth2013} and the  HUDF09-2
\citep{bouwens2011_hudf09}. 
A summary of all the data sets used in the present study is provided in Table \ref{tab:surveys}, along with the covered area and  the 5$\sigma$ depths. The latter 
are drawn from \cite{bouwens2015} and are based on the median
uncertainties in the total fluxes (after correction to total), as
found for the faintest 20\% of sources in the catalog.
As discussed by \cite{bouwens2015}, these depths reflect the actual sensitivity achieved in science images, as established through artificial source recovery simulations   \citep[see][for details]{bouwens2015}.

\begin{table}
\begin{center}
\caption{Datasets used}
\label{tab:surveys}
\begin{tiny}
\setlength{\tabcolsep}{1.5pt}
\begin{tabular}{l|c|ccccccccccccccc}
\hline
\hline
Field & Area   & \multicolumn{8}{|c}{5$\sigma$ depth} \\
	& (arcmin$^2)$ & \bb &\vv & \ii & \i814 & \zz & \yy & \jj & \hh \\
CANDELS GOODS &\multirow{2}{*}{64.5} &\multirow{2}{*}{27.7}&\multirow{2}{*}{28.0}&\multirow{2}{*}{27.5}&\multirow{2}{*}{28.0}&\multirow{2}{*}{27.3}&\multirow{2}{*}{27.5}&\multirow{2}{*}{27.8}&\multirow{2}{*}{27.5} \\
South Deep (GSd) & \\
CANDELS GOODS  & \multirow{2}{*}{60.9}&\multirow{2}{*}{27.5}&\multirow{2}{*}{27.7}&\multirow{2}{*}{27.2}&\multirow{2}{*}{27.0}&\multirow{2}{*}{27.2}&\multirow{2}{*}{26.7}&\multirow{2}{*}{26.8}&\multirow{2}{*}{26.7} \\
North Wide (GNw)  &\\
XDF & 4.7 & 29.6 &30.0 & 29.8 & 28.7 & 29.2 & 29.7 & 29.3 & 29.4\\
HUDF09-2 &4.7 & 28.3 & 29.3 & 28.8 & 28.3 & 28.8 & 28.6 & 28.9 & 28.7\\ 
\hline
\hline
\end{tabular}
\end{tiny}
 \tablecomments{Datasets used in the analysis along with area covered by each survey  and $5\sigma$ depth for the HST observations, obtained from \citet{bouwens2015}, based on median uncertainty in the flux measurements for faint sources.}
\end{center}
\end{table}

We exploit the data reduction and source catalog derived by
\citet{bouwens2015}.  Data were processed using the ACS GTO pipeline
APSIS \citep{blakeslee2003} and the WFC3/IR pipeline WFC3RED.PY
\citep{magee2011}, with final science imaging drizzled to a
0.$^{\prime\prime}$03-pixel scale. The photometric catalog has been
constructed using SourceExtractor \citep{bertinarnouts1996} after
PSF-matching imaging to the F160W filter. Multi-band photometric
information is available in the following optical bands: F435W, F606W,
F775W, F814W, F850LP (hereafter \bb, \vv, \ii, $I_{814}$, \zz,
respectively), as well as in the following near-IR bands: F098M,
F105W, F125W, F140W, F160W (hereafter $Y_{098}$, \yy, \jj, $JH_{140}$,
\hh, respectively.)  Complete details on data analysis and catalog
construction can be found in \cite{bouwens2015}.

To ensure robust results, we limit our analysis to sources with
detection in the \jj+\hh bands at high signal-to-noise ratio [${\rm
  S/N(JH_{\rm{det}})} > 6 $], defined as:

\begin{equation}
\frac{\rm S}{\rm N}=\frac{\rm FLUX}{\rm FLUXERR},
\end{equation}

\noindent \citep{stiavellibook} where FLUX and FLUXERR are the isophotal flux and its associated
error in the combined detection band, which we indicate as
$JH_{\rm{det}}$.\footnote{Note this is distinct from the F140W image,
  indicated as $JH_{140}$.} 
 We note that adopting an even higher S/N limit [${\rm
  S/N(JH_{\rm{det}})} > 8 $] samples would be even purer, but to the detriment
  of sample statistics.\footnote{We note that within uncertainties, applying a more stringent S/N cut 
  yields the same results, thus we opted for S/N$>$6 to include in the analysis a larger number of objects. }

In addition, with the goal of focusing on  contamination from
extended sources, we remove stellar-like sources, that is all sources with CLASSTAR$>$0.85 measured from the detection 
image [where SourceExtractor assigns CLASSTAR=0 to (very) extended sources and CLASSTAR=1 to point sources].
We  then proceed to select LBG sources at high redshift (or interlopers with similar colors at low redshift). 
We apply a color cut selection which is as
uniform as possible across across samples with different median redshifts, to ensure 
a consistency in the analysis. The adopted criteria can be summarized as follows. 

For $z\sim5$ candidates %(hereafter \vv-dropouts)
\begin{equation}
\begin{split}
V_{606}-i_{775}& >1.0 \\
z_{850}-H_{160}& <1.3 \\
V_{606}-i_{775}& >0.75(z_{850}-H_{160})+1.0
\end{split}
\end{equation}

For $z\sim6$ candidates %(hereafter \ii-dropouts)
\begin{equation}
\begin{split}
i_{775}-z_{850} & > 1.0\\
Y_{105}-H_{160}& < 1.0\\
i_{775}-z_{850}& > 0.75(Y_{105}-H_{160}) + 1.0
\end{split}
\end{equation}

For $z\sim7$ candidates %(hereafter \zz-dropouts)
\begin{equation}
\begin{split}
z_{850} - Y_{105} & > 1.0\\
J_{125} - H_{160} & < 0.45\\
z_{850} - Y_{105} & > 0.75(J_{125} - H_{160}) +1.0
\end{split}
\end{equation}

For $z\sim8$ candidates% (hereafter \yy-dropouts)
\begin{equation}\label{eq:cuts}
\begin{split}
Y_{105} - J_{125} & >1.0\\
J_{125} - H_{160} & <0.5\\
Y_{105} - J_{125} & >0.75(J_{125} - H_{160})+1.0.
\end{split}
\end{equation}

\noindent The color-color selection criteria listed above are not sufficient to
construct a sample of galaxies that are confidently at $z\gtrsim 5$
because intermediate redshift galaxies with a prominent spectral break
such as the 4000 \AA{} break may also fall in the color-color selection
regions typical of LBGs at higher redshift. 

Following the standard practice, we use the photometry in the bands
bluer than the putative Lyman break to separate high-$z$ sources,
which in the following we indicate as \emph{dropouts}, from
lower-redshift galaxies, which we label as
\emph{interlopers}. Specifically, 
$z\sim5$ dropouts (named as \vv-dropouts) are selected as sources with
S/N(\bb)$<$2, $z\sim6$ dropouts (named as \ii-dropouts)
with S/N(\bb)$<2$ and S/N(\vv)$<2$, $z\sim7$ dropouts (named as \zz-dropouts)
and $z\sim8$ dropouts (named as \yy dropouts) with S/N($x$)$<2$ and $\chi^2_{x} <3$, where
$\chi^2_{opt}$ is defined as
%%%%%%
\begin{equation}\label{eq:chi2}
\chi^2_{opt}=\sum_x\left[sgn(FLUX_x)\cdot\left(\frac{FLUX_x}{FLUXERR_x}\right)^2\right]
\end{equation}
%%%%%%%
In the equation FLUX$_x$ is the isophotal flux measured in a given band, FLUXERR$_x$ the uncertainty associated to the flux, 
and $x$ is intended to be  \bb, \vv, and \ii  bands for \zz-dropouts and \bb, \vv,  \ii and \i814  bands for \yy-dropouts \citep[see also][]{bouwens2011a}. %  for the \zz dropouts and  \bb, \vv,  \ii  and \i814 bands for \yy-dropouts.%
In addition, following \cite{bouwens2015}, \zz-dropouts are   selected as sources with  
S/N(\i814)$<$2, but \i814 is not used for computing the $\chi^2_{opt}$.

Finally, if a dropout satisfies more than one dropout selection, we
assign it to the highest redshift sample. This additional cut removes
 only a few percent of the sources (for example, in the GSd dataset,
 we identified 38 cases out of 870 dropouts).  In contrast, we do not apply this
restriction to interlopers, which thus may enter multiple selections. 
On average, at most 2 interlopers appear in two selections, and none appears at
the same time in all the samples. 

Finally, we highlight that the dropout sample may in general contain a
residual (small) fraction of low-$z$ galaxies that have not been
identified through the photometric analysis, because of lack of
sufficiently deep imaging in the blue. Hereafter, we call them
\emph{contaminants}. Interlopers and contaminants are the focus of our
investigation.

%%%%%%%%%%%
\begin{figure*}
\centering
\includegraphics[scale=0.35]{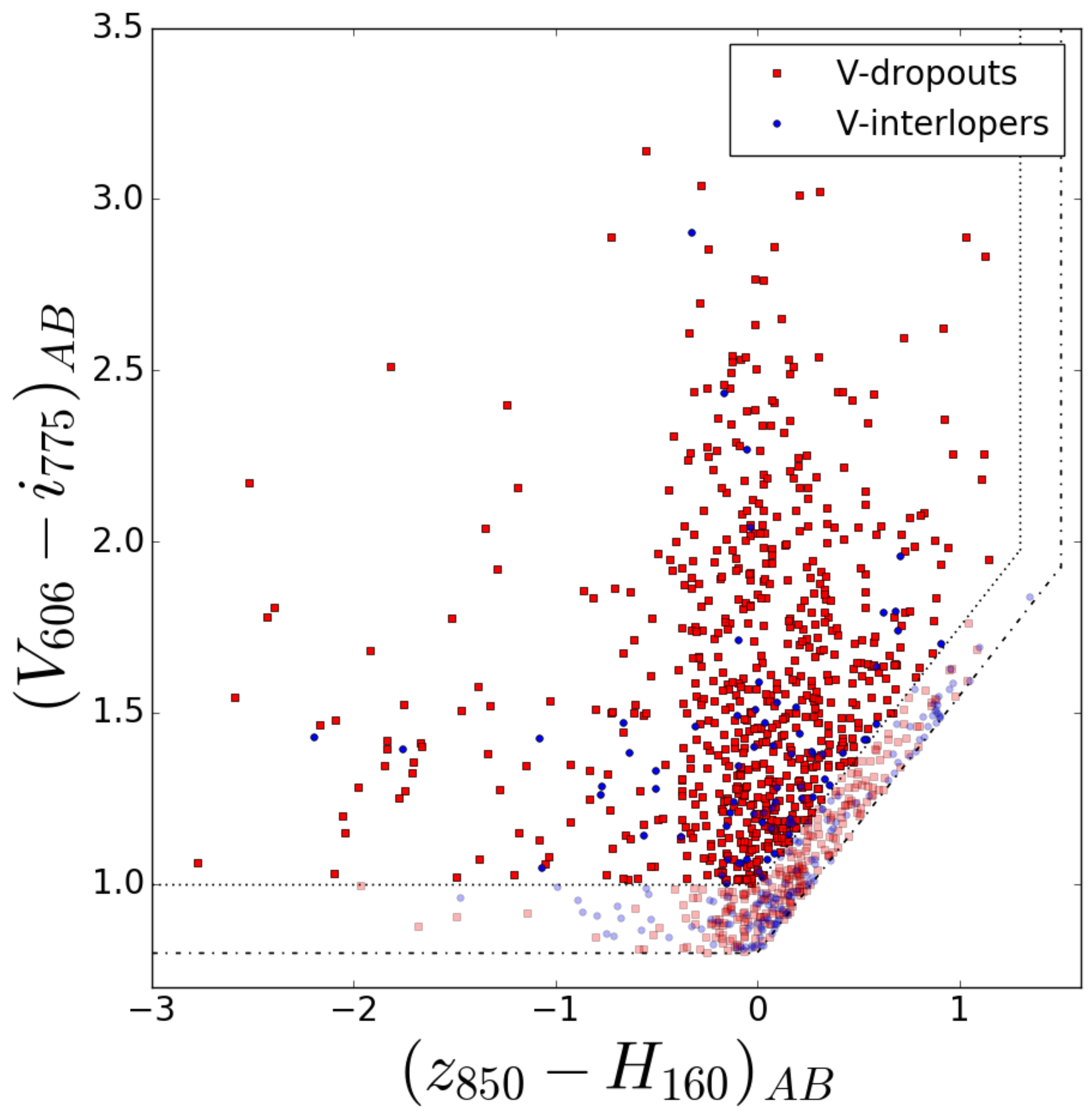}
\includegraphics[scale=0.35]{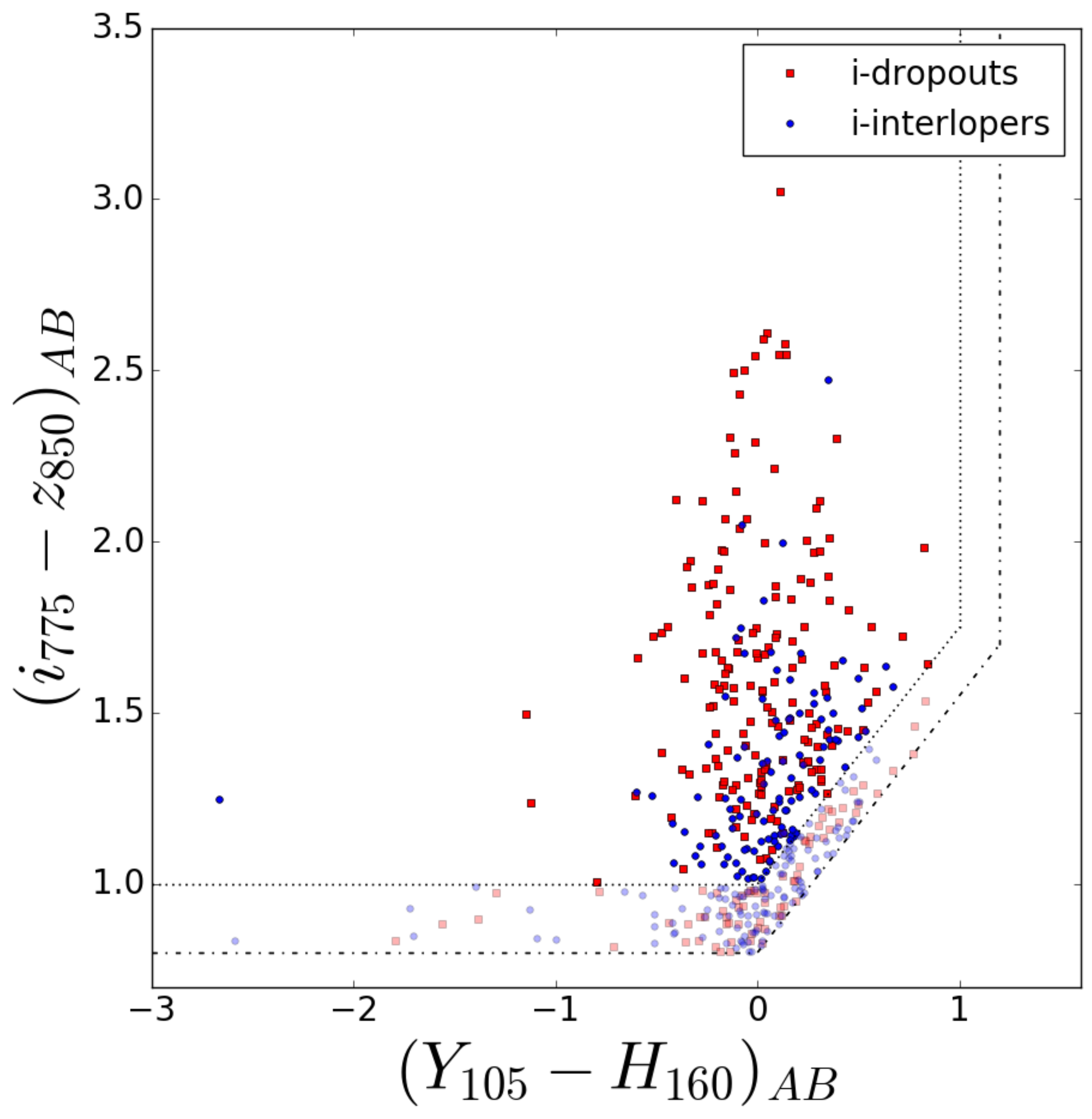}
\includegraphics[scale=0.35]{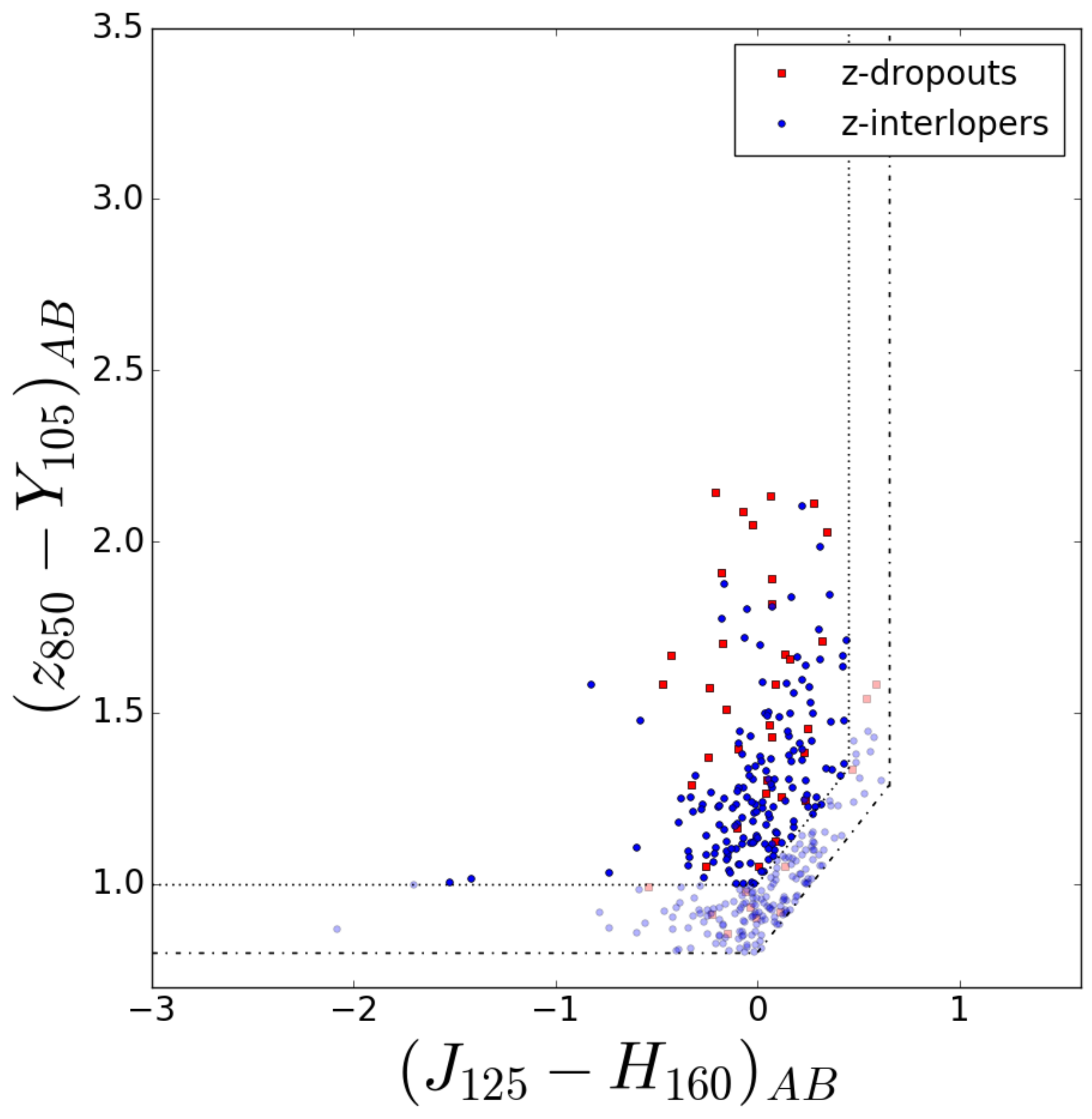}
\includegraphics[scale=0.35]{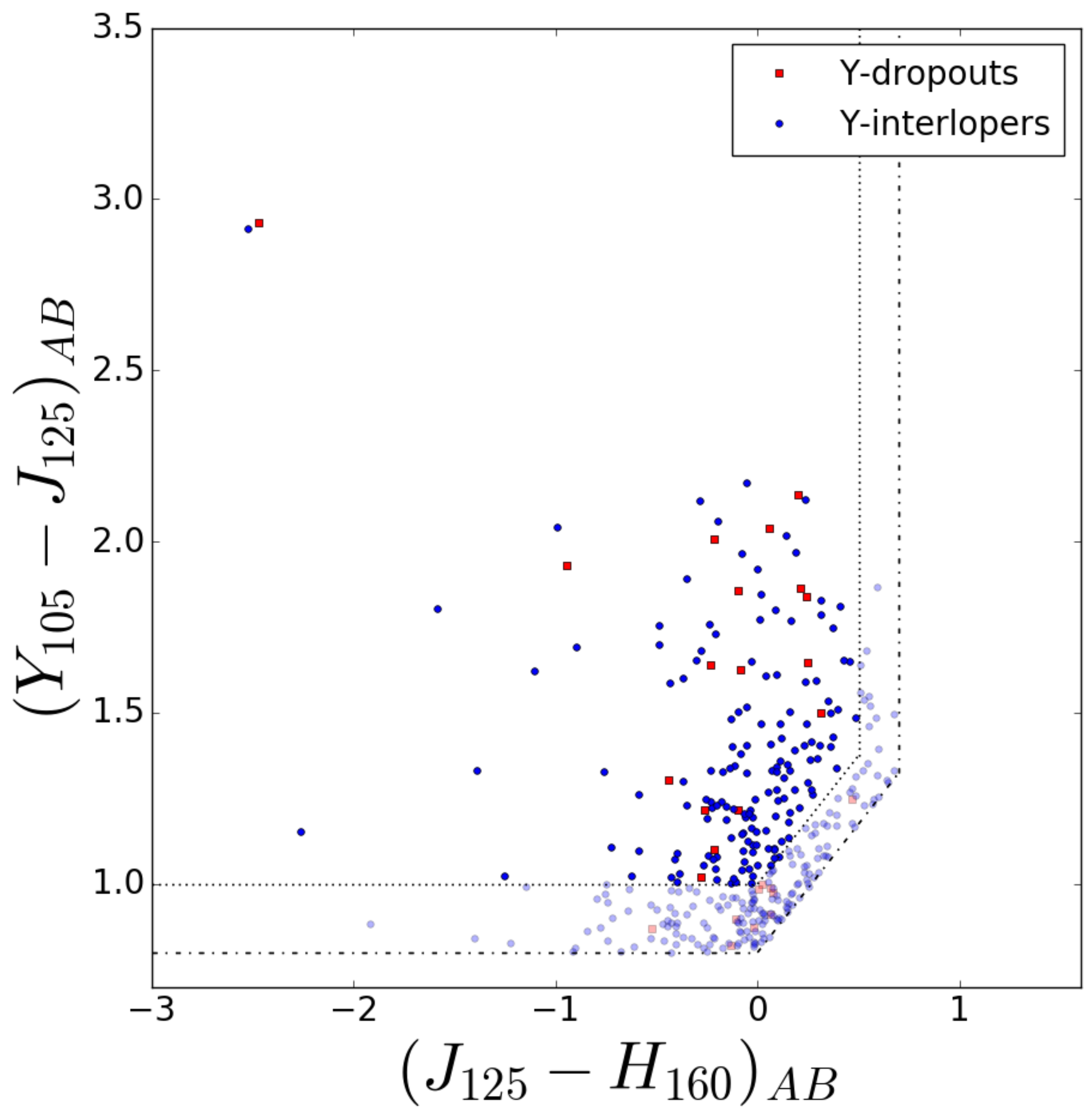}
\caption{Color-color selection box used to identify \vv- (upper left),
  \ii- (upper right), \zz- (bottom left) and \yy-dropouts (bottom
  right) over the GSd field. % over the GOODS-SOUTH DEEP field. %(upper panel) and SHALLOW (bottom panel), respectively. For each sample,
  Red squares represent dropouts, i.e high-$z$ sources with no flux
  blueward of the Lyman-break; blue circles represent interlopers,
  i.e.  high-$z$ candidates showing a detection in the blue bands.
  Dashed lines represent the boundaries of the original sample
  selection; dash-dotted lines represent
  the boundaries of the enlarged sample (see text for details). Darker
  symbols refer to the original selection, lighter ones to the
  enlarged
  selection.} %Stars are indicated with a starry symbol. \bbv{stars still need to be excluded from the sample!}}
\label{Fig:selection_box}
\end{figure*}
%%%%%%%%%%

Note that the separation between dropouts and
contaminants for sources with a low $\chi^2_{opt}$ is arbitrary to a
certain extent; for example, \cite{bouwens2015} impose a cut at
S/N$<2$, while the Brightest of Reionizing Galaxies survey (BoRG,
\citealt{trenti2011}) resorts to a more conservative threshold of
S/N$<1.5$ in the bluest bands. Obviously,  more conservative cuts 
entail the exclusion of a higher number of  real high$-z$ sources from the selections, 
thus different investigators may decide to prioritize differently sample purity versus selection completeness.

\section{Results}\label{Sec:characterization_contaminants}

%%%%%%%%%%%%%%%%%%%%%%%
\begin{table*}
\begin{center}
\caption{Statistics of dropouts and interlopers }
\label{tab:dropouts_percentages}
\begin{tiny}
\setlength{\tabcolsep}{3pt}
\begin{tabular}{l|cc|cc|cc|cc|cc|cc|cc|cc}
\hline
\hline
 & \multicolumn{4}{c|}{GSd}  & \multicolumn{4}{c|}{GNw}& \multicolumn{4}{c|}{XDF}  & \multicolumn{4}{c}{HUDF09-2}\\

\multirow{2}{*}{population} & \multicolumn{2}{c|}{original sample}  & \multicolumn{2}{c|}{enlarged sample}& \multicolumn{2}{c|}{original sample}  & \multicolumn{2}{c|}{enlarged sample}&\multicolumn{2}{c|}{original sample}  & \multicolumn{2}{c|}{enlarged sample}& \multicolumn{2}{c|}{original sample}  & \multicolumn{2}{c}{enlarged sample}\\
     		& number & $\%$ & number & $\%$ & number & $\%$ & number & $\%$ & number & $\%$ & number & $\%$ & number & $\%$ & number & $\%$ \\    
\hline
\vv-dropouts	& 648 & 90$\pm$2	& 882&81$\pm$2 	& 392 &87$\pm$2	& 510&73$\pm$2	&132 &93$\pm$3	&165&85$\pm$4 	&102&94$\pm$4 	&127&88$\pm$4\\
\vv-interlopers	& 72 &10$\pm$2 	& 205 & 19$\pm$2 	& 58& 13$\pm$2	& 189 &27$\pm$2	&10 &7$\pm$3 		&28&15$\pm$4 	&6 &6$\pm$4		&17&12$\pm$4\\
\hline
\ii-dropouts	& 172 & 62$\pm$4 	& 239&52$\pm$3 	&72 & 65$\pm$7 	&105&47$\pm$4	&69&77$\pm$7		&78&68$\pm$6 	&28&60$\pm$10 	& 37&51$\pm$8\\
\ii-interlopers 	& 106 &38$\pm$4 	& 223 & 48$\pm$3 	&39&35$\pm$7 	& 118&53$\pm$4 	&20&23$\pm$7		&37&32$\pm$6 	&16& 40$\pm$10	& 35&49$\pm$8\\
\hline
\zz-dropouts 	& 33 & 17$\pm$4 	& 42&11$\pm$2 	&29 & 26$\pm$6 	&35&16$\pm$4 	&31&40$\pm$8 	&36&30$\pm$6		&17& 30$\pm$10 	&22&23$\pm$6\\
\zz-interlopers 	& 157 &83$\pm$4 	& 322& 89$\pm$2 	&82 & 74$\pm$6 	&180&84$\pm$4 	&47&60$\pm$8		&85&70$\pm$6 	&37 &70$\pm$10	&73&76$\pm$6\\
\hline
\yy-dropouts 	& 17 & 10$\pm$3 	& 27&8$\pm$2 		&54 &36$\pm$6 	&62&26$\pm$4 	&6 &50$\pm$20	&10&40$\pm$20	&12 &21$\pm$8	&15&17$\pm$6\\
\yy-interlopers 	& 150 &90$\pm$3 	& 329 & 92$\pm$2 	&96 &64$\pm$6 	&177&74$\pm$4 	&6&50$\pm$20		&15&60$\pm$20 	&45 &79$\pm$8 	&71&83$\pm$6\\
\hline
\hline
\end{tabular}
\end{tiny}
 \tablecomments{ Errors are defined as binomial errors \citep{gehrels86}.}
\end{center}
\end{table*}
%%%%%%%%%%%%%%%%%%%%

In this section we present and discuss our results obtained separately for the four field we analyze (GSd, GNW, XDF, HUDF09-2). However, we  only show plots 
for GSd, to avoid unnecessary repetitions.

% \bbv{say it better}
\subsection{Numbers and redshift distribution of dropouts and interlopers}

The color-color selection of dropouts and interlopers is shown in
Figure \ref{Fig:selection_box} for samples of \vv, \ii, \zz, and  \yy-dropout
sources drawn from the GSd field, with discrimination between the two classes based on the 
S/N and 
optical $\chi^2$ (Equation~\ref{eq:chi2}). %It is immediately apparent
We note that %A large fraction of sources are located close to the boundaries of the
%selection box, and 
photometric scatter is likely to play a
significant role in the selection of faint objects. Indeed, more than half of the 1$\sigma$ error bars for the interlopers
intersect at least one boundary of the color-color selection box. Therefore, to
carry out a more comprehensive analysis, we enlarge the color-color
selection box by 0.2 mag, to check for both candidate high-$z$ LBGs
and interlopers that slightly fail to meet the adopted selection
criteria (see also \citealt{su2011} for an alternative approach based
on assigning a probability that a source belongs to the color-color
selection). In the following, we will call \emph{original} selection
the one given in \S\ref{Sec:dataset}  (dotted line in Fig.
\ref{Fig:selection_box}), and \emph{enlarged} selection the one
introduced in this section (dash-dotted line in Fig.
\ref{Fig:selection_box}).

%While the number of interlopers seems fairly constant across the different
%redshift bins, that of dropouts does not. Therefore, 
The most striking feature of Figure~\ref{Fig:selection_box} is the
relative weight of interlopers vs. dropouts, which is quantified in
Table~\ref{tab:dropouts_percentages} for all the fields considered. We first focus on the GSd field. 
At  lower redshift
($z\sim 5$), dropouts dominate the sample within the original selection,
while the opposite situation is present at higher redshift
($z\sim 8$), when the interloper fraction is much higher. 
We stress that this percentage  is not giving a level of contamination 
in our dropout sample, since the presence of sufficiently deep data at bluer wavelengths allows us to identify  the interlopers.
Interestingly, the situation remains qualitatively similar
with our enlarged selection, although as expected the enlarged samples
contain a larger fraction of interlopers. If we adopted 
a more conservative S/N in the sample selection (S/N$<$1.5), percentages
of dropouts would be systematically smaller, but comparable within 2$\sigma$ 
uncertainty. 

The observed behaviour is mainly due to the fact that %t two effects. First, if we examine 
the population of
dropouts %we notice a 
steady decreases in  number %of sources 
for the
higher redshift selections. This is primarily determined by the
evolution of the luminosity function, which decreases significantly
from $z\sim 5$ to $z\sim 8$ at all luminosities. %Second, 
In contrast, the number
of interlopers in the sky remains approximately constant over a wide range of dropout selection windows.
%has the opposite trend, and there is an increase
%of the population over the same area of sky with increasing redshift of the
%Lyman-break selection. 
Second order effects in the evolution of the interloper
population with the redshift of the Lyman break selection are 
complex to model, and include intrinsic evolution of their luminosity
functions, change in the distance modulus and in the comoving volume
of the selection, with partial offsets among them (e.g., the decrease
in sensitivity because of an increase in the distance modulus is
offset by an increase of the comoving volume). 

Similar findings are obtained when we analyze the other fields, even
though the results from the different fields highlight the presence
of sample (or ``cosmic'') variance, naturally expected because of
galaxy clustering (see, e.g., \citealt{trenti2008}).  In addition, fields
such as the XDF and HUDF09-2 have small areas, resulting in
significant Poisson uncertainty. 
Finally, the difference in relative depths reached by the different surveys  plays a
role, which we discuss further in Section \ref{sec:contamMC}.

%%%%%%%%%%%%%%%%%%%%%%%%
\begin{figure}
\centering 
\includegraphics[scale=0.45]{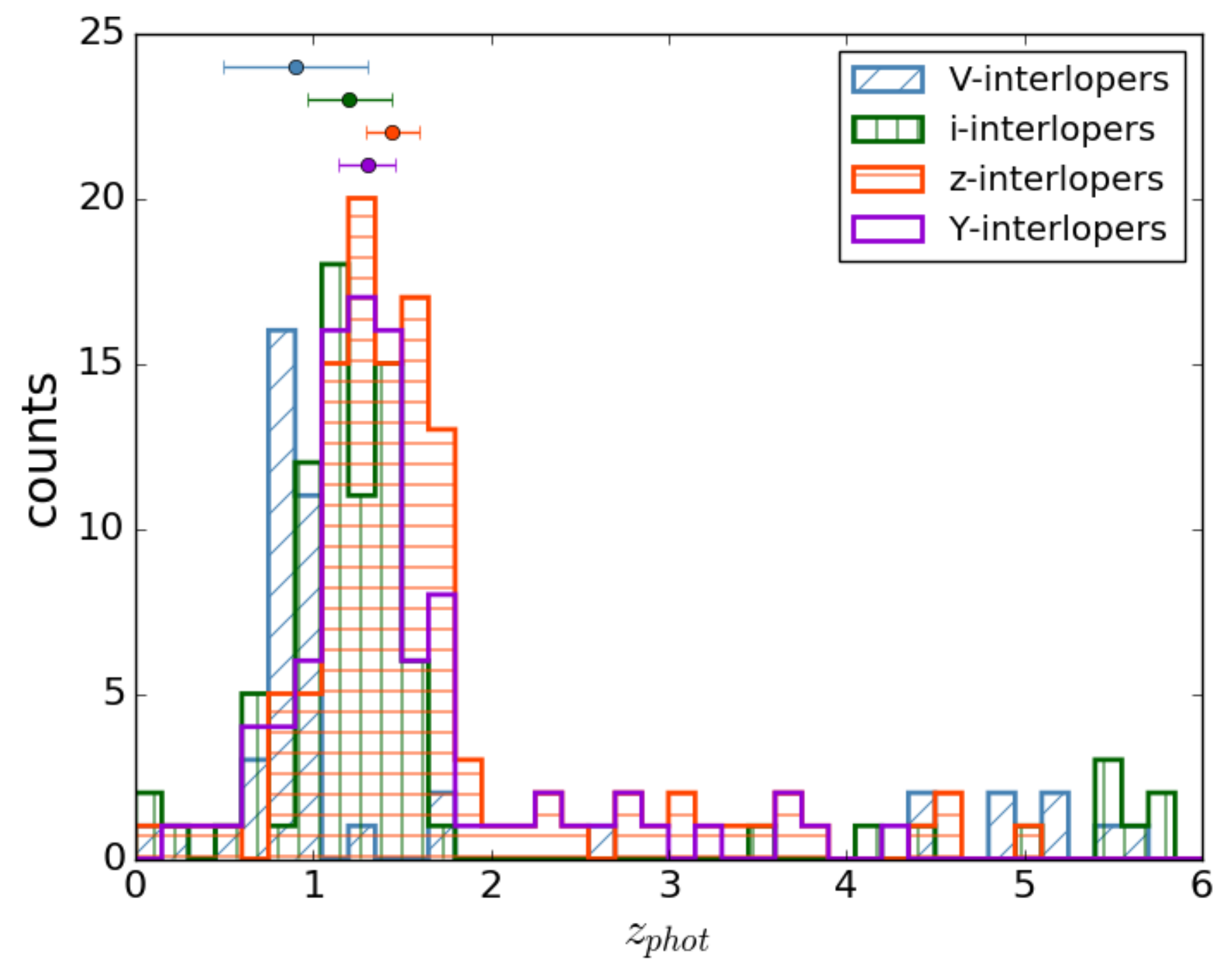}
\caption{Redshift distribution for the interlopers at the different
  dropout selections, as indicated in the labels, for the original
  sample over the GSd field. Photometric redshifts are drawn from the
  3D-\emph{HST} survey \citep{skelton2014}. Median values along with
  their associated uncertainty (defined as 1.235$\times\sigma/\sqrt{n}$, with $n$
  number of objects) are shown above the histograms as horizontal points with error bars.}
\label{Fig:zphot}
\end{figure}
%%%%%%%%%%%%%%%%%%%%%%%%%

To further investigate the properties of the interlopers, and test
whether these are indeed 4000 \AA{} break sources, we resort to the
photometric redshift catalogs from the 3D-\emph{HST} survey
\citep{skelton2014}, which we matched to our sources based on
coordinates and \hh band magnitudes. The expectation is that given
$z_{dropouts}$ as the redshift of the Lyman-break selection, the
interlopers should be peaked at $z_{interlopers}$ given by:
%%%%%%
\begin{equation}\label{eq:zint}
 1+ z_{interlopers}=\frac{1216}{4000}\times (z_{dropouts}+1).
\end{equation}
%%%%%
So, for example, for z = 5 selection, the interlopers are predicted to be found at
$z\sim 0.7-1.0$ corresponding to the Balmer and 4000\AA{} breaks; 
for z=8 selection, the interlopers are expected at $z\sim 1.6-1.9$. 
%So, for example, for dropouts at $z\sim5$, interlopers are expected at
%$z\sim0.7$; for dropouts at z$\sim8$ interlopers are expected to have
%$z\sim1.8$. 
Taking into account uncertainties, this is broadly the case based on the photo-$z$ analysis,
as shown in Figure \ref{Fig:zphot} for the GSd field. For this field, after the match with the 3D-\emph{HST} survey, we recover $\sim85\%$
of our sources.  From this figure, and from
Equation~\ref{eq:zint}, it is clear that as $z_{dropouts}$ increases,
$\langle z_{interlopers} \rangle$ changes relatively little ($\Delta (z)<1$).  
The error on the median values narrows as we go from lower to higher redshift, 
but this is mainly due to the larger sample statistics provided by the \yy-interlopers
with respect to the \vv-interlopers.  
The fact that not all interlopers of a given selection fall exactly in
the expected redshift window, and the lower than expected median
redshift of the Y-interlopers
highlight the limitations of both our selection method and photo-z techniques.
Indeed, some real dropouts might be misclassified due to photometric scatter
and/or the photo-z estimates might not be reliable.
Similar trends are obtained also for the other
fields, even though uncertainties are very large. 

Thus, extrapolating the
trend to even higher redshift samples of dropouts, such as those
accessible by JWST observations, one expects that the number of
interlopers in the color-color selection will remain relatively
constant, while dropout numbers will decrease very rapidly for
$z_{dropouts}>10$ based on theoretical modeling
(see, e.g., \citealt{mason2015}). 

We note that, according to the Madau-Lilly plot
(e.g. \citealt{madau14}), the star formation rate peaks at
intermediate redshift ($z\sim 2$). This means that the number density
of interlopers for $z_{interlopers}\gtrsim1.85$ (corresponding to
$z_{dropouts}\sim9.5$ from Equation~\ref{eq:zint}) may likely slightly
decrease with increasing redshift, although the decrease of the
interloper density will still be less steep than the one of the
dropouts because the latter have significantly higher redshift.

\subsection{Surface densities of dropouts and interlopers}

%%%%%%%%%%%%%%%%%
\begin{figure}[!t]
\centering
\includegraphics[scale=0.2]{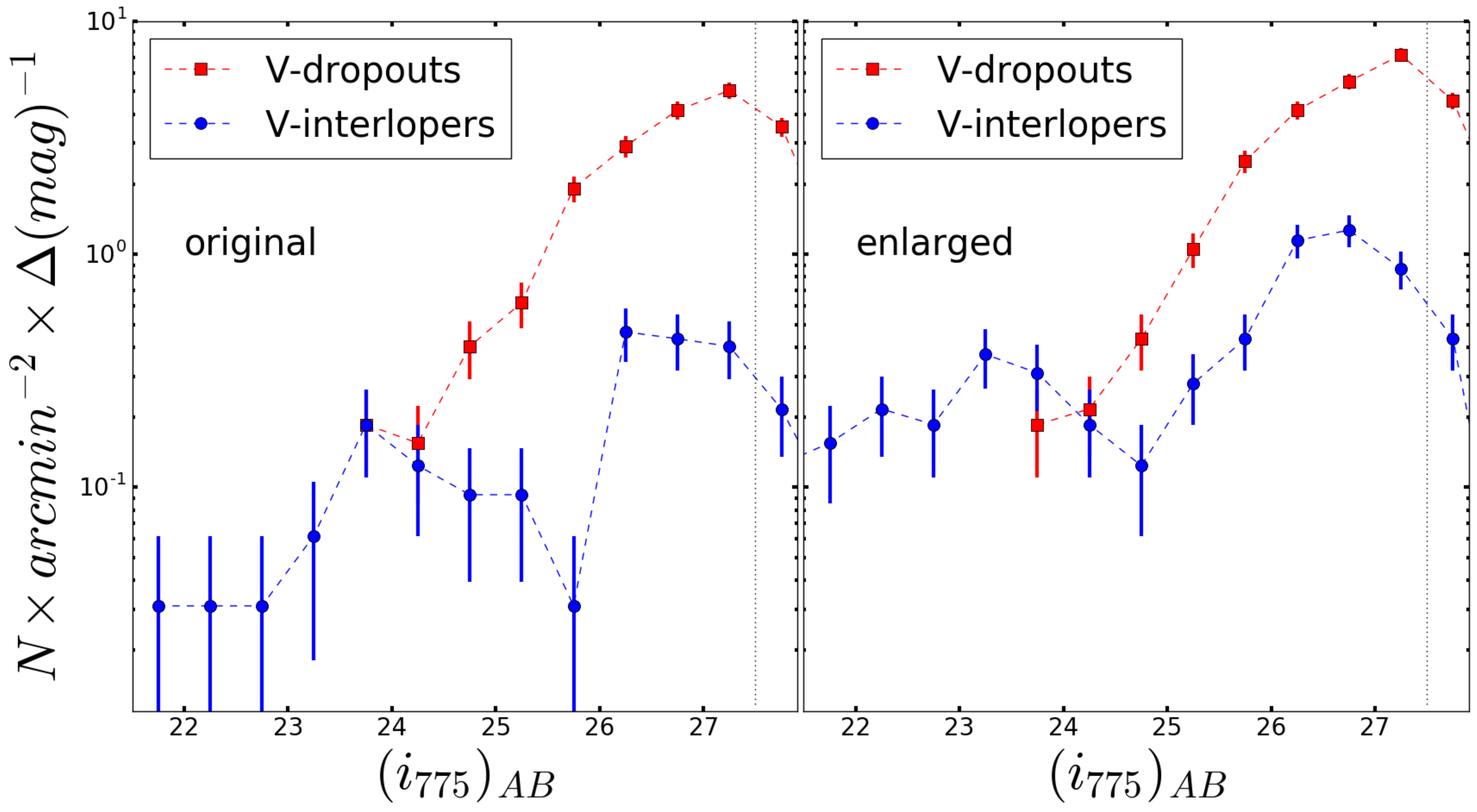}
\includegraphics[scale=0.2]{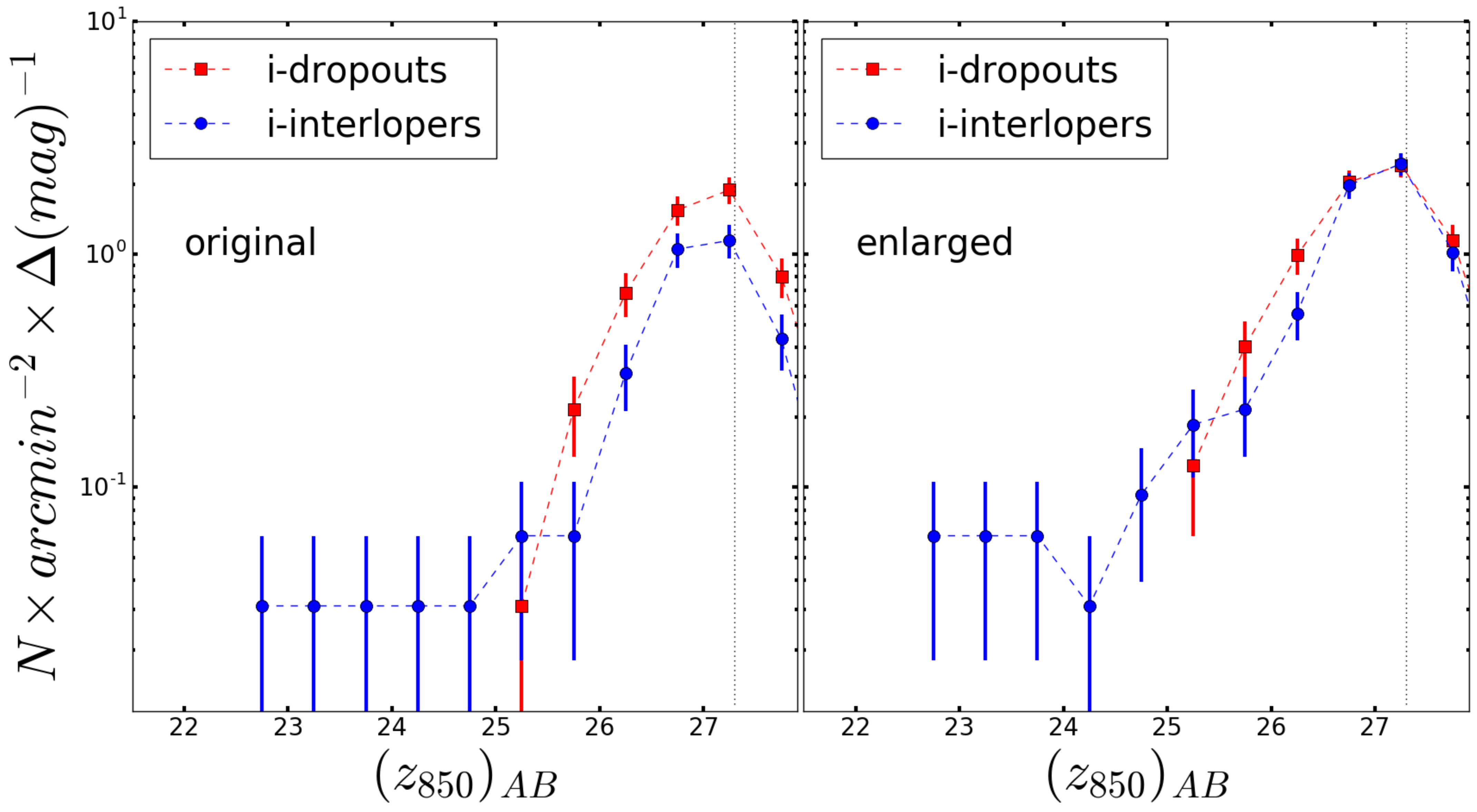}
\includegraphics[scale=0.2]{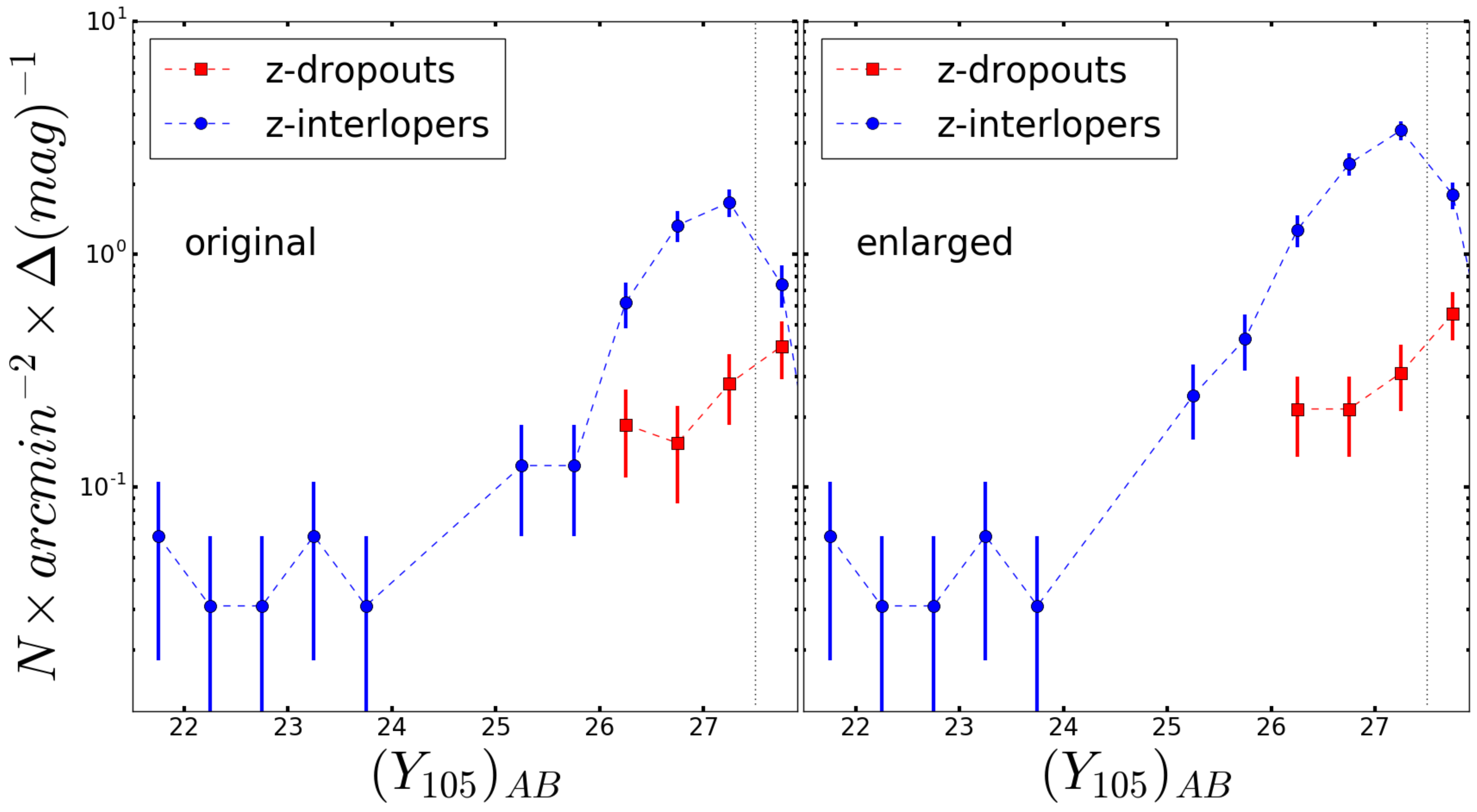}
\includegraphics[scale=0.2]{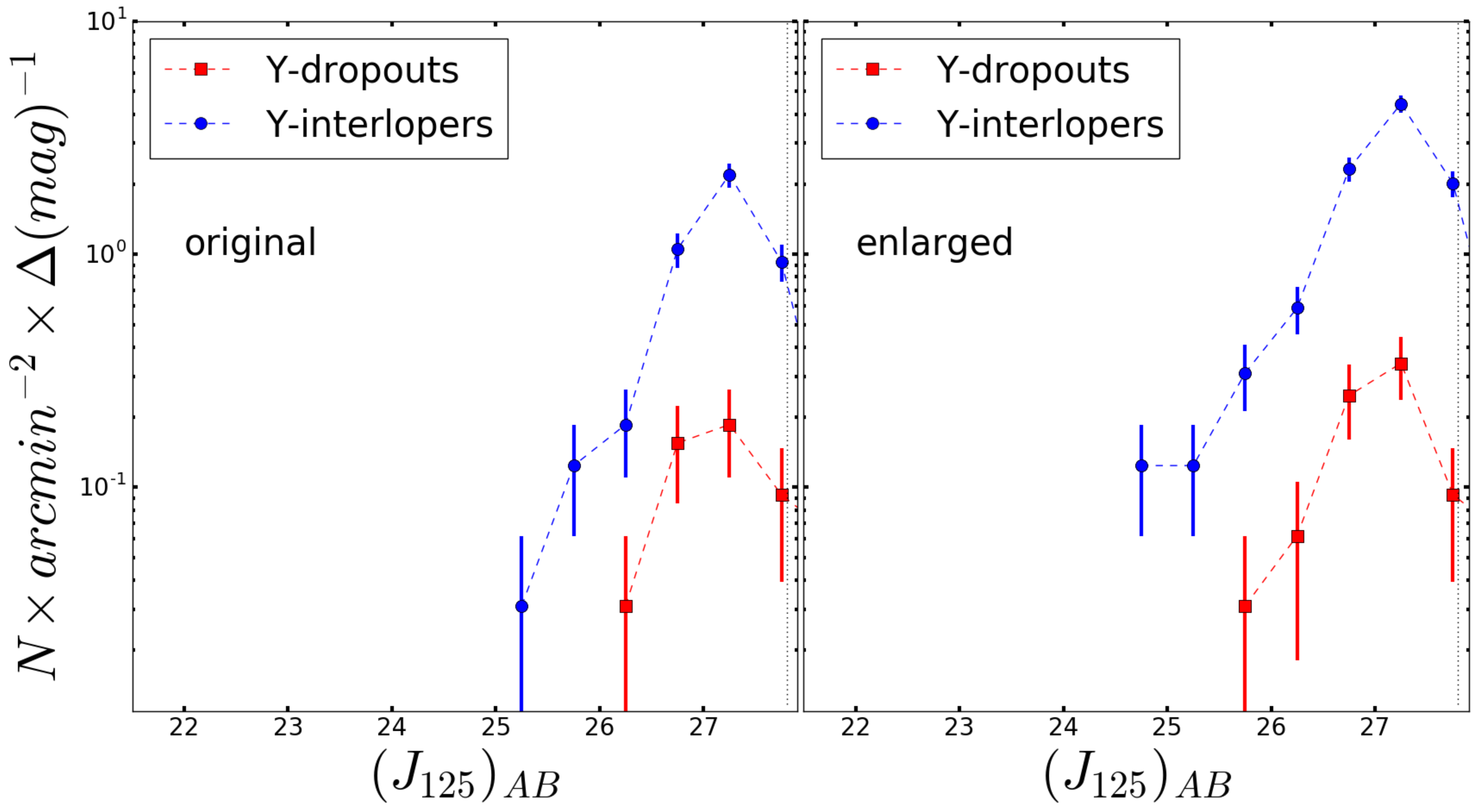}
\caption{Surface density distribution  of dropouts (red) and interlopers (blue) in the original (left) and enlarged (right) samples for the selections at the different redshifts, as indicated in the labels, for the GSd field. %Red and blue horizontal lines indicate the magnitude range over which the fit has been performed, for dropouts and interlopers, respectively. 
The black vertical line indicates the formal 5$\sigma$ magnitude limit \citep{bouwens2015}.} %In the labels, numbers represent the number of objects in each sample. }
\label{Fig:surf}
\end{figure}
%%%%%%%%%%%%%%%%%%

In the previous section we have investigated the incidence
of dropouts and interlopers at the different redshifts. We now aim at characterizing the distribution of
luminosity for these populations, to study how they compare. % of interlopers is, and how it compares to
%that of dropouts. 
Thus, we derive the surface density distributions of
dropouts and interlopers by counting the number of objects in each bin
of 0.5 magnitudes and dividing it by the area of the survey, %(64.5
%arcmin$^2$), 
as shown in Figure \ref{Fig:surf} for the GSd field. For each population,
the surface density is plotted as a function of \ii for the $z\sim5$
samples, \zz for the $z\sim6$ samples, \yy for the $z\sim7$ samples
and \jj for the $z\sim8$ samples. These are the magnitudes in the band
that best matches the 1600  \AA\/ rest-frame at that redshift for the
dropouts, as done in \cite{bouwens2015}. We note that for  $m_{AB}\gtrsim 27$ (see the exact value for each magnitude as
dotted line in Fig.  \ref{Fig:surf}) all our samples suffer from
incompleteness, which is the cause of the apparent decline in the number
counts of faint objects. 

Surface density distributions strongly depend on the redshift and on
the population considered.  At the lowest redshift, the surface
density distribution of interlopers is relatively flat with magnitude
for $20\leq$ \ii $\leq 28$ and there are about 0.1 interlopers per
$arcmin^{2}$ in bins of magnitude. In contrast, the distribution of
dropouts rises very steeply. As expected due to the well established
exponential cut off of the luminosity function at the bright end
\citep[e.g.][]{bouwens2015, bowler2014, oesch2014, mclure2013, schenker2013}, there are
essentially no dropouts brighter than \ii$\sim$24. Overall, dropouts
are much more numerous than interlopers. Similar conclusions are
reached in both the original and enlarged samples.

Moving to higher redshift, the shape of the distribution of dropouts
stays almost constant, just showing a modest steepening, but that of
interlopers considerably changes. At $z\sim6$, interlopers and dropouts have similar distributions, with the
exception that interlopers extend toward brighter magnitudes. At $z\sim7$ and
$z\sim8$ interlopers are more numerous than dropouts at all
luminosities, and have a tail of objects at the bright-end as well.
Overall, the interloper distribution appears as steep as the 
dropout one. This holds both for the original and the enlarged
samples. %, with the exception that in the enlarged sample a larger
%number of bright interlopers enters the selection.
Similar results are found also in the other fields.

%%%%%%%%%%%%%%%%%
\begin{figure}[!t]
\centering
\includegraphics[scale=0.2]{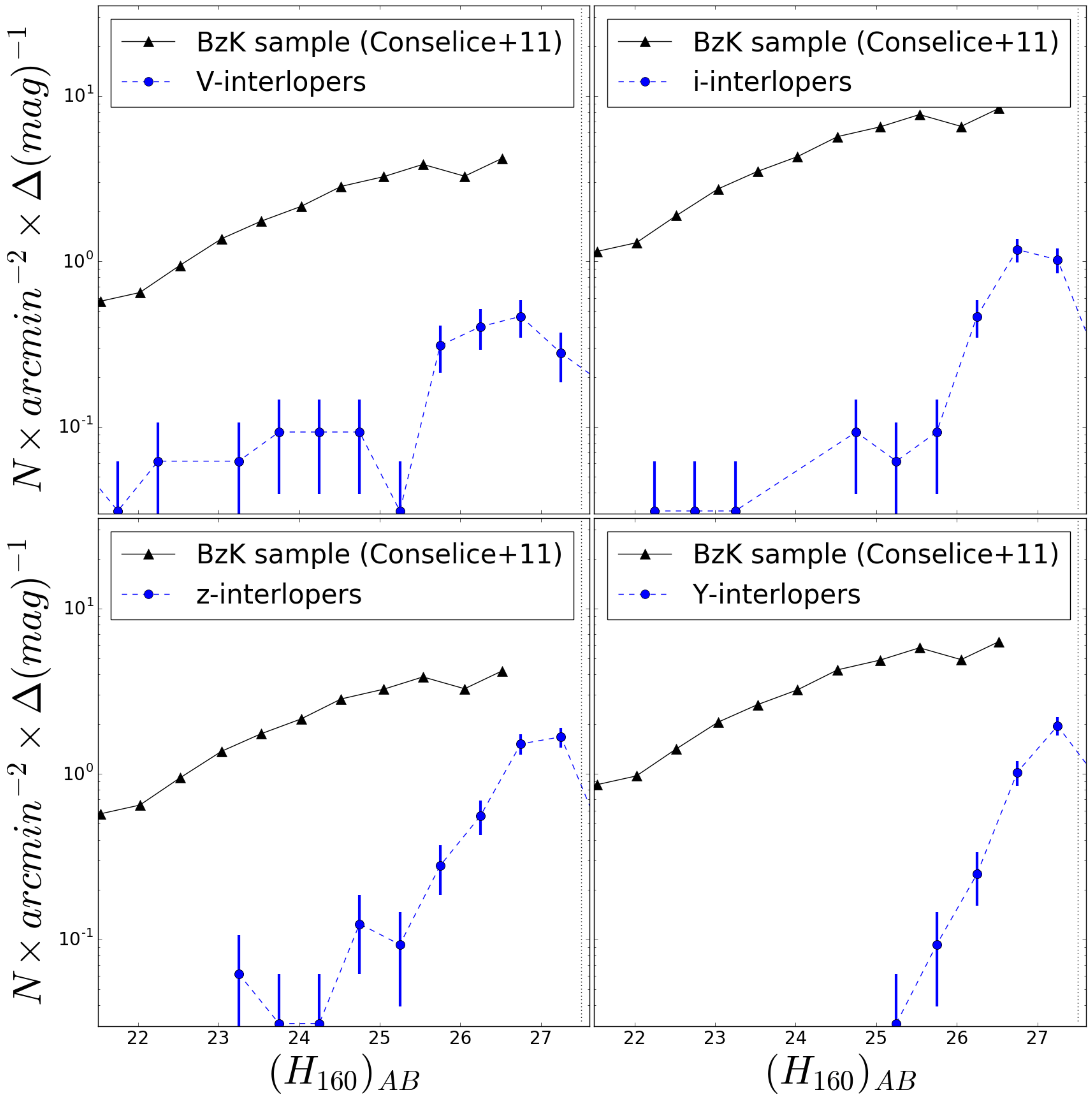}
\caption{$H_{160}$ surface density distribution  of  interlopers (blue) in the original sample for the selections at the different redshifts, as indicated in the labels, for the GSd field. Samples of BzK selected samples, drawn from \citep{conselice2011}, are shown from comparison. %Red and blue horizontal lines indicate the magnitude range over which the fit has been performed, for dropouts and interlopers, respectively. 
The black vertical line indicates the formal 5$\sigma$ magnitude limit \citep{bouwens2015}.} %In the labels, numbers represent the number of objects in each sample. }
\label{Fig:surfH}
\end{figure}
%%%%%%%%%%%%%%%%%%

It is reasonable to expect that interlopers are a subpopulation of  BzK color- selected galaxies \citep{daddi2004, daddi2007}.  
This method is  designed to find red, dusty or passively evolving older galaxies at $z > 1.5$. 
We can therefore compare our derived surface densities of interlopers to those of BzK selected samples. 
We use as reference the dataset presented by \cite{conselice2011} for  galaxies at $1.5<z<3$ 
drawn from GOODS North and South fields and the GOODS NICMOS Survey. 
That study analyses two of the same fields considered in our work and
includes HST imaging, 
therefore reaching a deeper magnitude limit compared to the 
many studies of BzK samples conducted from the ground (e.g. Cirasuolo et al. 2007, Hartley et al. 2008).  
\cite{conselice2011} quote the \hh-magnitude distribution of all galaxies at $1.5<z<3$, without splitting them into redshift bins, 
so a direct comparison to our results is not possible because our
interloper samples are more localized in redshift (see Fig.\ref{Fig:zphot}). Still, to have a
first-order approximation, we treat the \cite{conselice2011}
sample as uniform in redshift, and thus we simply rescale the observed
number counts to take into consideration the difference in volume
with our selections. 

Figure \ref{Fig:surfH} compares the \cite{conselice2011} scaled distribution 
to the \hh-band number counts for our interlopers samples in the GSd field.  
At each magnitude and in each redshift bin, the BzK population is up to a factor of 10 larger than that of  interlopers. 
This is consistent with the assumption that not all galaxies at $z\sim1.5-2$ are interlopers of high redshift selections, 
but only the subset with a particular combination of
colors. Interestingly, we observe that the interloper counts get
steeper at faint luminosities with increasing 
redshift compared to the general BzK sample. This might suggest that
interlopers evolve differently compared to the general population, but
investigating this trend in more details is beyond the scope of this
work.

\subsection{Distribution of optical $\chi^2$ for interlopers}

%%%%%%%%%%%%%%%%%%%%%%%%
\begin{figure}
\centering
\includegraphics[scale=0.35]{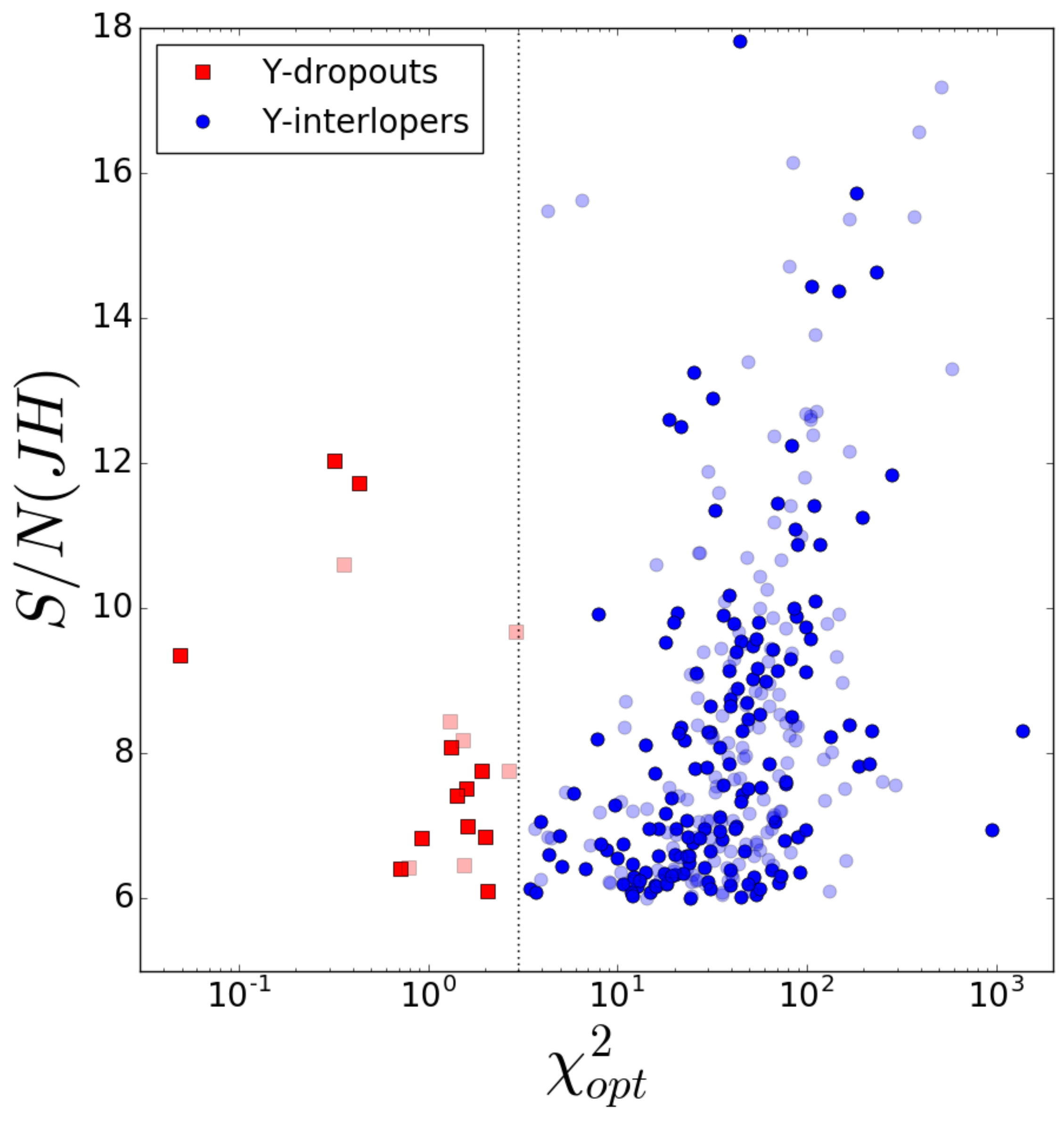}
\caption{Comparison between the ${\rm S/N(JH_{\rm{det}})}$ and the optical
  $\chi^2$ for \yy-dropouts for the GSd field. Colors and symbols are as in
  Fig. \ref{Fig:selection_box}. The dotted vertical line
  indicates the separation between interlopers and dropouts adopted in
  this work.}
\label{Fig:SN}
\end{figure}
%%%%%%%%%%%%%%%%%%%%%%%%%

One of the aims of our analysis is to  derive an estimate of the contamination
in dropout samples. Before proceeding, we need to characterize 
the distribution of the optical $\chi^2$ values for
interlopers as a function of their S/N in the detection bands. It is
evident that the robustness of the detection is a key quantity to
distinguish between interlopers and dropouts. In fact, while less deep
observations in the redder bands give smaller ${\rm S/N(JH_{\rm{det}})}$ and
simply exclude galaxies both from the dropout and interloper
populations, less deep observations in the bluer bands produce lower
$\chi^2_{opt}$ and may induce a mis-classification, moving sources
from the interloper to the dropout population.

Figure \ref{Fig:SN} plots the ${\rm S/N(JH_{\rm{det}})}$ to the $\chi^2_{opt}$
for both dropouts and interlopers from the $z\sim 8$ selection (\yy
Lyman-break selection), for the GSd field. As expected, the interlopers have a
positive correlation  between the two quantities, reflecting the finite
amplitude of the 4000  \AA{} break. Similar results also hold for samples from
selections at lower $z$ and drawn from the other fields.

\subsection{Contamination in dropout samples}\label{sec:contam}

Now that we characterized the properties of interlopers, we can use
them to investigate dropout sample contamination  induced by
interlopers that are mis-classified as dropouts in absence  of sufficiently 
deep data at bluer wavelengths. The main causes of
contamination are the impact of noise in the measurement of the optical
$\chi^2$ and photometric scatter in the color-color selection.

%%%%%%%%%%%%%%%%%%%%%%%
\begin{table*}
\begin{center}
\caption{Statistics of dropouts and interlopers after the MCMC experiment that introduces spurious noise on the observations}
\label{tab:dropouts_percentages_SN}
\begin{tiny}
\setlength{\tabcolsep}{3pt}
\begin{tabular}{l|cc|cc|cc|cc}
\hline
\hline
 & \multicolumn{2}{c|}{GSd}  & \multicolumn{2}{c|}{GNw}& \multicolumn{2}{c|}{XDF}  & \multicolumn{2}{c}{HUDF09-2}\\

\multirow{2}{*}{population} &original sample  &enlarged sample& original sample  &enlarged sample&original sample  &enlarged sample&original sample  &enlarged sample\\
     		 & $\%$  & $\%$ & $\%$ & $\%$ & $\%$ &  $\%$ &  $\%$ & $\%$ \\    
\hline
\vv-dropouts & 72$\pm$2 & 69$\pm$2 & 71$\pm$3 & 66$\pm$3 & 75$\pm$6 & 72$\pm$5 &81$\pm$6 & 78$\pm$6\\
\vv-interlopers & 28$\pm$2 &31$\pm$2 & 29$\pm$3 &34$\pm$3 & 25$\pm$6 &28$\pm$5 & 19$\pm$6 &22$\pm$6\\
\hline
\ii-dropouts &32$\pm$3 & 30$\pm$3 &34$\pm$6 &30$\pm$4 &52$\pm$8 &47$\pm$7 &33$\pm$9 &29$\pm$7\\
\ii-interlopers &68$\pm$3 & 70$\pm$3 & 66$\pm$6 &60$\pm$4 &48$\pm$8 &53$\pm$7 & 67$\pm$9 &71$\pm$7\\
\hline
\zz-dropouts &48$\pm$2 & 4$\pm$1 & 9$\pm$3 &7$\pm$2 &9$\pm$5 & 8$\pm$4 &8$\pm$5 &7$\pm$4\\
\zz-interlopers &52$\pm$2 &96$\pm$1 & 91$\pm$3 &93$\pm$2 &91$\pm$5 & 92$\pm$4 & 92$\pm$5 &93$\pm$4\\
\hline
\yy-dropouts &2$\pm$1 & 2.2$\pm$0.1 & 4$\pm$2 &3$\pm$1 &4$\pm$7 & 5$\pm$5 & 5$\pm$4 &4$\pm$3\\
\yy-interlopers & 98$\pm$1 & 97.8$\pm$0.1 & 96$\pm$2 &97$\pm$1 &96$\pm$7 &95$\pm$5 & 95$\pm$4 & 96$\pm$3\\
\hline
\hline
\end{tabular}
\end{tiny}
 \tablecomments{ Errors are defined as binomial errors \citep{gehrels86}.}
\end{center}
\end{table*}
%%%%%%%%%%%%%%%%%%%%

To estimate the impact of noise in the measurements on the datasets we analyzed, we perform a
resampling Monte Carlo (MC) simulation on the entire photometric
catalogs and  add zero-mean noise in the fluxes sampling from a
Gaussian distribution with width determined by the S/N ratio of the
simulated broadband fluxes. We then apply the dropout
selection criteria given in Sec.\ref{Sec:dataset}, and quantify the
number of interlopers and dropouts in the simulated sample. 
 We repeat the procedure 500 times to
collect statistics and we find that on average increasing the noise we obtain  systematically larger
fractions of interlopers at any redshift than those obtained with the original catalogs (Tab.\ref{tab:dropouts_percentages}).
The average statistics are given in Tab.\ref{tab:dropouts_percentages_SN} for each field separately.
This test emphasizes the need of precise photometry to robustly distinguish between dropouts and interlopers. 

We note that if instead of using the entire catalogs as starting point of the MC experiment
we used only a combination of the dropout and interloper  enlarged samples, we would get results in agreement 
within the errors, indicating that actually only the sources close to the boundaries of our selection boxes
can contaminate the samples.

As the next step, we also consider the photometric scatter and 
perform a more sophisticated
resampling MC simulation on the photometric
catalogs. Specifically, for each dropout selection, we uniformly
sample with repetition the luminosity in the detection band from the
catalog of enlarged interlopers, extracting a simulated catalog with
the same size as the original one. Next, we assign to each of these
objects the broadband colors of a random galaxy from the same
catalog (again using uniform sampling probability with
repetition), and we add zero-mean noise in the fluxes sampling from a
Gaussian distribution with width determined by the S/N ratio of the
simulated broadband fluxes. We use as our starting point a catalog that includes
all the \yy-interlopers detected in our four fields (enlarged samples), 
in order to consider a population that is relatively homogenous, but statistically significant. Note that for this second test 
it would not be appropriate to resort to the photometric catalogs of all sources, 
since the MC procedure effectively ``re-shuffles'' colors of galaxies, thus a relatively uniform starting sample is needed.  
Finally, we perform the photometric
analysis of the catalog to quantify the number of interlopers in the enlarged sample that are
classified as dropouts. After repeating the procedure 500 times to
collect statistics, for example for the GSd field we find that on average:
%%%%
\begin{itemize}
\item The $z\sim 5$ selection has 17$\pm$4 interlopers entering the
  \vv-dropout sample as contaminants, for an estimated contamination rate
  $f_c\sim 17/648 \sim 2.6\%$;
\item The $z\sim 6$ selection has 7$\pm$2 interlopers entering the 
  \ii-dropout sample as contaminants, for an estimated contamination rate 
  $f_c\sim 7/172 \sim 4.0\%$;
\item The $z\sim 7$ selection has 2$\pm$1 interlopers entering the 
  \zz-dropout sample as contaminants, for an estimated contamination rate 
  $f_c\sim 2/33 \sim 6.0\%$;
\item The $z\sim 8$ selection has 1$\pm$1 interlopers entering the 
  \yy-dropout sample as contaminants, for an estimated contamination rate 
  $f_c\sim 1/17 \sim 5.9\%$.
\end{itemize}

%%%%%
%Note that the
%  error on the number of contaminant only takes into account the
%  statistical uncertainty on the 500 realizations, while a larger,
%  systematic contribution is given by the intrinsically small number
%  of objects in the enlarged interloper sample (see Table~\ref{tab:dropouts_percentages}).
  
%\bbv{compare these with contamination estimates from Bouwens et al. (2015) as well (maybe mentioning that the exact procedure is simulated later)}
Overall, results from the different fields are in agreement,  indicating that the contamination
is always only a few percent in all samples, and it increases with increasing 
redshift. These results are also in broad agreement with other literature estimates, as it will be discussed in 
Sec.\ref{Sec:lit}. 

These results are clearly illustrating that while the number of
mis-classified interlopers remains relatively constant across
different samples, as the redshift increases, the relative weight
compared to the number of dropouts grows significantly. Interestingly,
these estimates are consistent with the predictions from the
contamination model based on source simulations from an extensive SED
library encompassing a wide range of star formation histories,
metallicities, and dust content and a combination of an old and a
young population \citep{oesch2007}, and used for the BoRG survey
sample purity analysis (e.g., see
\citealt{trenti2011,bradley2012,calvi2016}).  %Figure \ref{Fig:conts}
%shows the redshift distribution for \vv, \ii, \zz, \yy-dropouts as
%derived from 
Applying the color cuts adopted in the current work, the model predicts a contamination of 0.7\% at
$z\sim5$, 1.6\% at $z \sim6$, 3.5\% at $z\sim7$ and 7.3\% at
$z\sim8$. 

%%%%%%%%%%%%%%%%%%%%%%

\subsubsection{Contamination at $z\sim8$}\label{sec:contamMC}
%%%%%%%%%%%%%%%%%%%%%%%%
\begin{figure}
\centering
\includegraphics[scale=0.3]{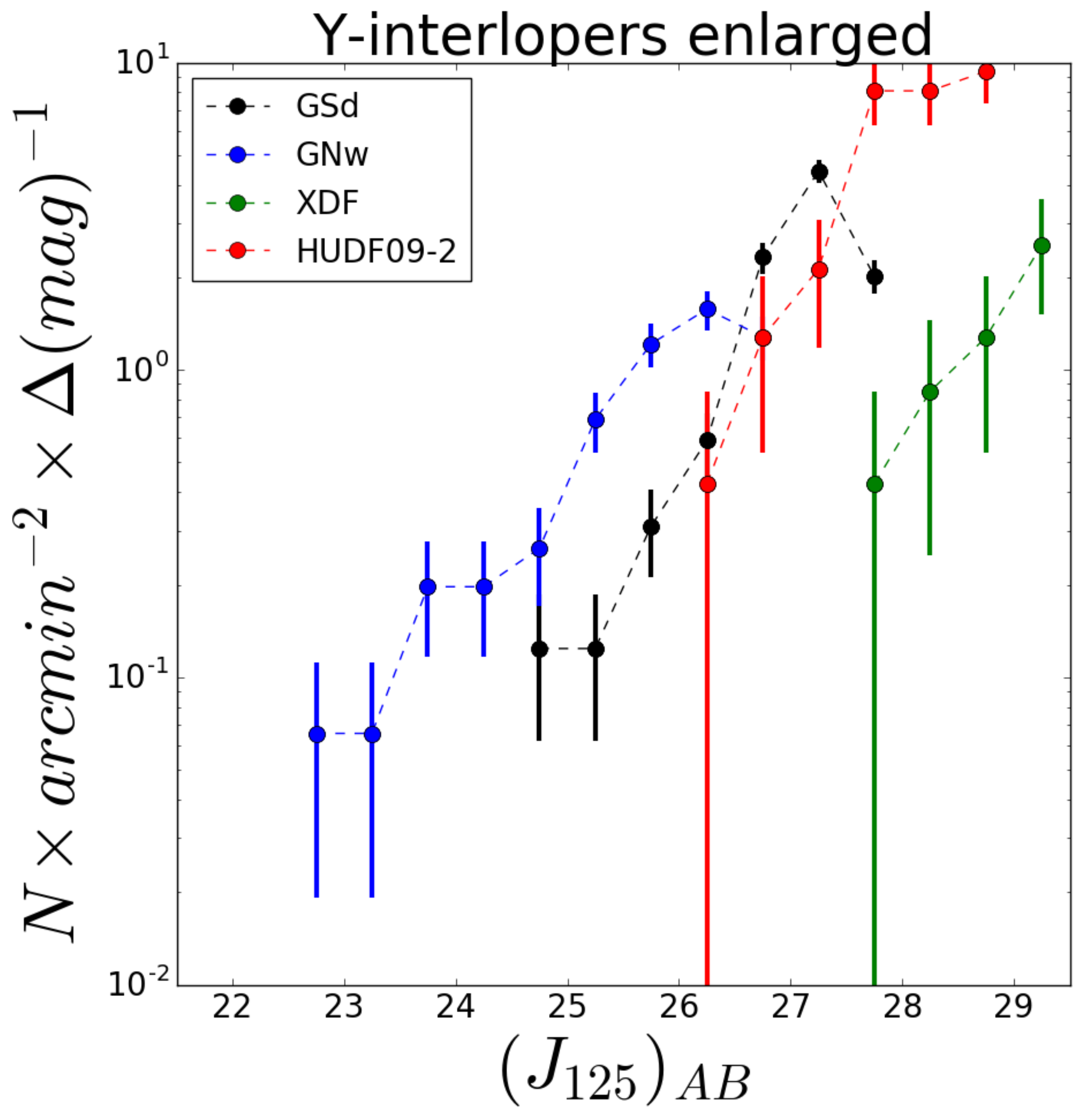}
\caption{Surface density distribution  of \yy-interlopers in the enlarged  samples  for all the fields considered in this study, as indicated in the labels. Each sample is plotted down to its  formal 5$\sigma$ magnitude limit \citep[from][]{bouwens2015}. 
The best power law fit of the sample is $\log(N/arcmin^2/\Delta(mag)) = (0.35\pm0.1)\times$ \jj $+ (-9.0\pm0.4)$.\label{Fig:SB_all}}
\end{figure}
%%%%%%%%%%%%%%%%%%%%%%%%%

%%%%%%%%%%%%%%%%%%%%%
\begin{figure*}
\centering
\includegraphics[scale=0.4]{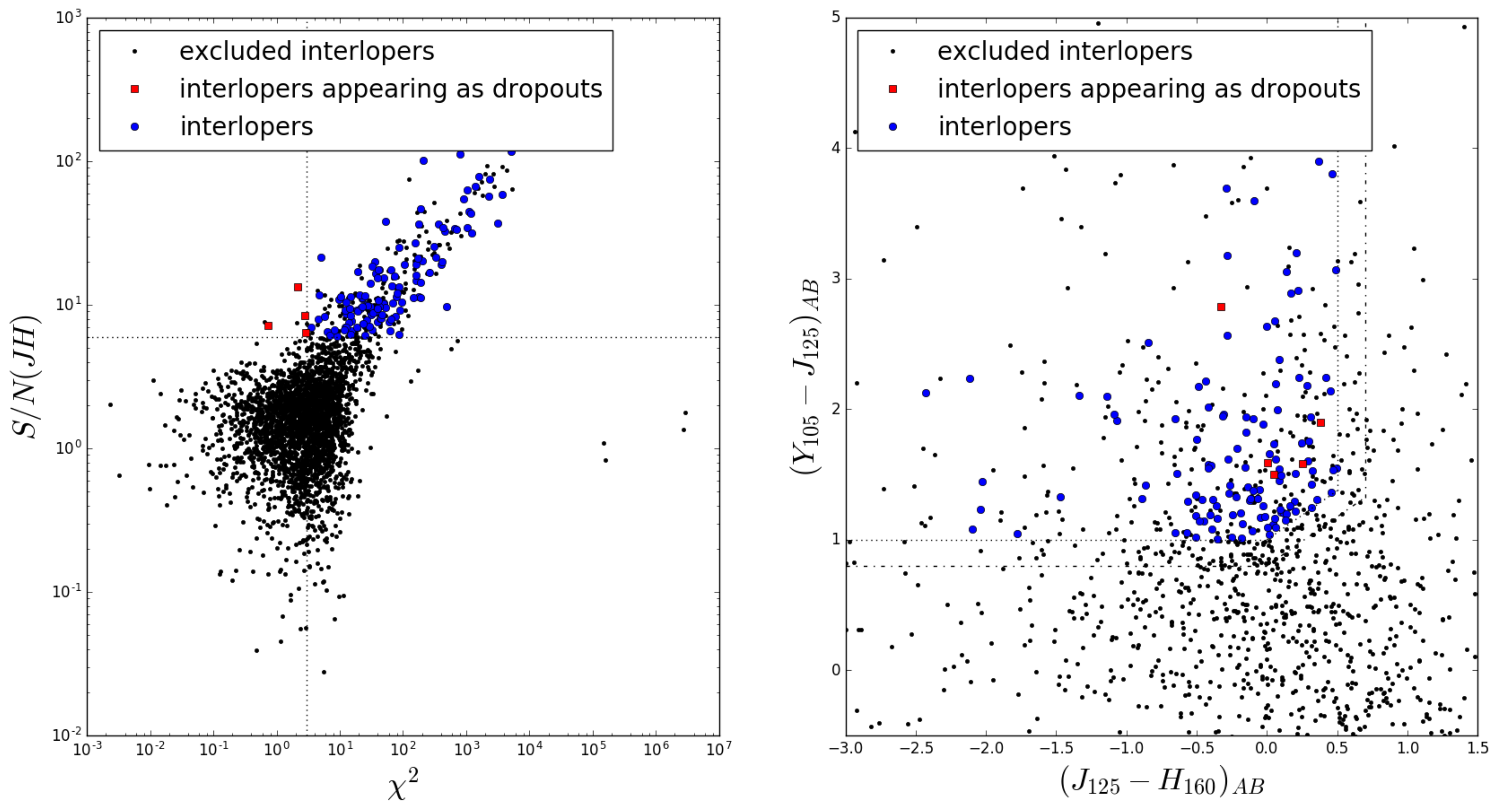}
\caption{Results of one realization of our MonteCarlo extraction aimed at testing the reliability of the interloper-dropout selection in deeper surveys (\jj$\leq$30) for a XDF-like survey (see text for details). Left: comparison between the ${\rm S/N(JH_{\rm{det}})}$ and the optical $\chi^2$. Dotted lines represent the  detection thresholds. Right: color-color selection box used
  to identify \yy-dropouts. Blue squares represent interlopers, red squares dropouts and black dots galaxies that do not enter anymore the selection after the dimming procedure.  Dotted and dash-dotted lines are as in Fig.  \ref{Fig:selection_box}.} %In the labels, numbers represent the number of objects in each sample. }
\label{Fig:MC}
\end{figure*}
%%%%%%%%%%%%%%%%%%%%%%
%%%%%%%%%%%%%%%%%%%%%%%%% 
 \begin{figure}
\centering
\includegraphics[scale=0.45]{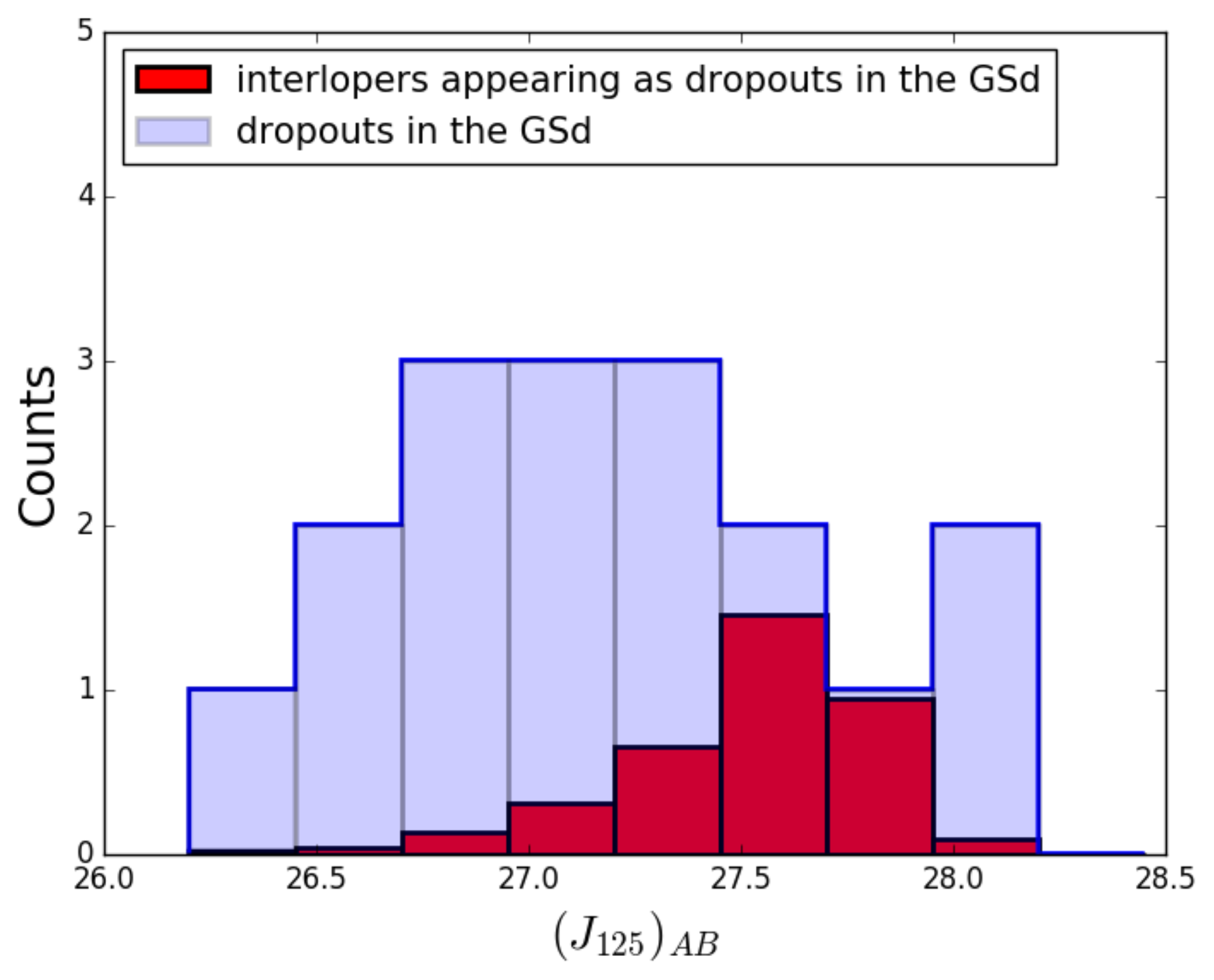}
\caption{ \jj magnitude distribution for the sample of \yy-dropouts in the GSd as derived from the cuts in Sec.\ref{Sec:dataset} (blue)  and for the average contamination from dropouts as obtained from our Monte Carlo simulation (red) (see text for details).  } %In the labels, numbers represent the number of objects in each sample. }
\label{Fig:mag_interlopers}
\end{figure}
%%%%%%%%%%%%%%%%%%%%%%%%%

We now focus only  on the
highest redshift selection ($z\sim 8$), since it contains the largest
number of interlopers, and investigate more in detail the level of contamination 
in the different surveys. 

Using a similar approach to that presented in the previous section, we can use the multi-band photometric
catalog of all the interlopers identified in the different fields (enlarged sample), 
combined with an extrapolation of
the surface number densities of interlopers, in order to estimate the
contamination in different surveys, assuming that the spectral energy
distributions of the interlopers  are
representative of the general population. We simulate a series of surveys with
relative depths in the different bands similar to those used in our analysis, but different values of 
5$\sigma$ depth.

For the simulation, we compute the brightness distribution of  all
the \yy-interlopers in the enlarged sample (Fig.\ref{Fig:SB_all}).
For a basic characterization of the luminosity distribution, we fit
the populations using a power law through a Markov Chain Monte Carlo
method. The best power law fit of the sample is
$\log(N/arcmin^2/\Delta(mag)) = (0.35\pm0.1)\times$ \jj
$+ (-9.0\pm0.4)$.  As shown in Fig.\ref{Fig:SB_all}, while trends for
the GSd and HUDF09-2 are compatible within the errors, the GNw field
seems to have a  systematically  larger number of interlopers, while the
XDF has a systematically lower number. This plots confirms that there
is significant cosmic variance across fields. Quite interestingly, GNw
not only has an excess of interlopers, but there is also an excess of
genuine high redshift candidates reported by many studies
\cite{finkelstein2013, oesch2014, bouwens2015} and across a range of
redshifts. Further medium/deep lines of sights beyond those available
in the HST archive would be very interesting to investigate in greater
detail this correlation. 

We then assume that all interlopers in the sky follow a power-law fit
to the number counts distribution, extrapolated in the magnitude range
\jj=22-31, and we sample from this distribution a catalog of object
luminosities. Next, we proceed to estimate the contamination in each
field separately. We assign to each simulated object the broadband
colors of a random galaxy from one of our interloper sample (GSd, GSw,
XDF, HUDF-092), and add to the signal in each band a Gaussian noise
drawn from a distribution with dispersion associated to the S/N that
would be achieved in the simulated survey (following the depths
reported in Tab.\ref{tab:surveys}). Finally, we apply the dropout
selection criteria given in Sec.\ref{Sec:dataset}, and quantify the
number of simulated interlopers that satisfy the dropout
selection. This gives us our best estimate of the surface number
counts of contaminants in each simulated survey.

For the GSd simulation, based on the extrapolated number density of
interlopers for 22$\leq$\jj$\leq$31 and an assumed area of $64.5$
arcmin$^2$, we extract $4390$ simulated galaxies per realization, and
we repeated the simulation 500 times.  On average, we find that a
realization has 4$\pm$2 of these simulated interlopers appearing
as dropouts, 120$\pm$10 are correctly classified as interlopers, while
the remaining either turn out to have ${\rm S/N(JH_{\rm{det}})}<6$, or colors
outside the selection box, and do not enter in the interloper or
dropout sample. To illustrate the MC experiment, results are shown in
Fig. \ref{Fig:MC} for one of the 500 realizations. This implies that the
probability of
mis-classifying a specific (enlarged sample) interloper as dropout is
very small $p\sim 0.001$.

Figure \ref{Fig:mag_interlopers} shows the \jj magnitude distribution
for observed dropouts in the GSd, selected  with the criteria
given in Sec.\ref{Sec:dataset}. Overplotted is also the average magnitude
distribution of the simulated contaminants (i.e. interlopers appearing
as dropouts after the MC experiment). It can be clearly seen that the
fraction of contamination increases at fainter magnitudes, consistent
with the explanation that photometric scatter is the main cause of
contamination. 

Repeating the MC experiment for the other fields, we found that for
the GNw/XDF/HUDF09-2 simulation, on average, a realization has
1$\pm$1/1$\pm$1/1$\pm$1 of the simulated interlopers appearing as
dropouts, and 53$\pm$7/45$\pm$6/25$\pm$5 that are correctly classified
as interlopers.  This implies that the probability of mis-classifying
a specific (enlarged sample) interloper as dropout is very small in
all the fields ($p\sim 0.0002/0.003/0.003$).  The GNw field is the one
with the lowest contamination. As shown in Tab.\ref{tab:surveys}, this
field has the deepest relative depth blueward of the dropout band
compared to the detection band, and clearly this allows more efficient
identification of interlopers, minimizing contamination of the dropout
sample.  In fact, even though the other fields have deeper photometry,
their photometric limits in the blue bands are relatively shallower
compared to the limits of the detection (red) bands, inducing a higher
probability of misclassification of interlopers as dropouts (and
therefore higher contamination).

Overall, our analysis also suggests that given a survey, the higher the S/N
in the detection, the higher the likelihood that the dropout is a
LBG. Thus we are fully consistent with
the high sample purity inferred from spectroscopic follow-up studies
of LBG samples in ultradeep surveys
(e.g. \citealt{malhotra2005}) since spectroscopic investigations are
limited to the brighter galaxies (e.g.  $m\lesssim 27.5$).

%%%%%%%%%%%%%
\begin{figure}
\centering
\includegraphics[scale=0.45]{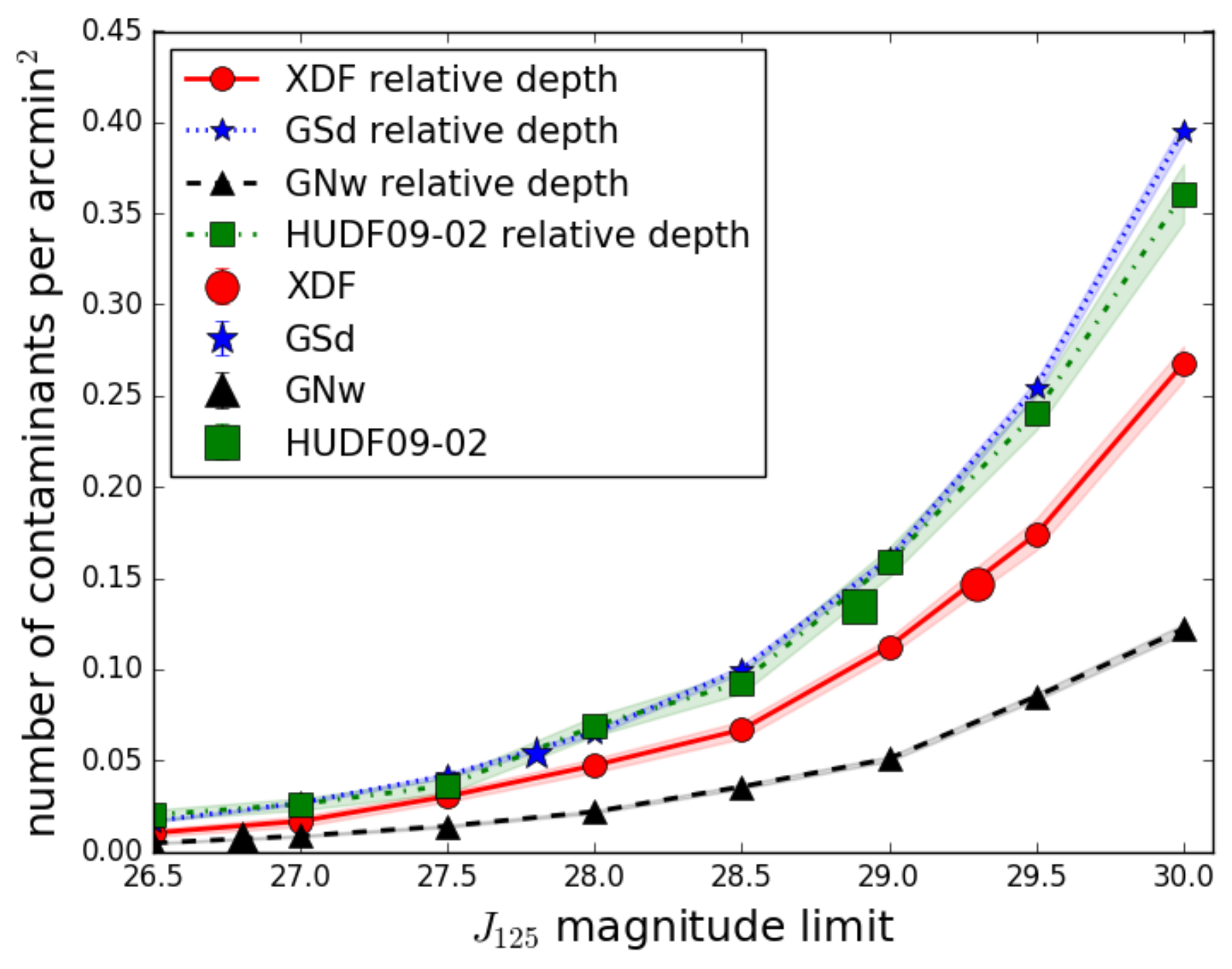}
\caption{Estimated level of contamination in \yy-dropout surveys due
  to misidentified interlopers depending on survey magnitude
  limits. Different colors refer to different observational choices
  (relative depths and filter types) used a reference, as indicated in
  the label. Errors are the errors on the mean of the 500
  realizations.
} %In the labels, numbers represent the number of objects in each sample. }
\label{Fig:cont}
\end{figure}
%%%%%%%%%%%%%

To explore how contamination changes as a function of the limiting
depth of the survey, we repeat the MC experiments varying the \jj band
magnitude limit and scaling the limits in all other bands keeping the
relative depths constants %(simulating the three cases XDF, HUDF09-1,
%and HUDF09-2). 
The results are shown in Figure \ref{Fig:cont} which
summarizes the level of contamination per arcmin$^2$ versus limiting
magnitude. As expected, the contamination increases toward fainter
magnitudes, because there is a higher number of potential contaminants,
 and strongly depends on the relative limiting depths of the
different bands. Overall, the level of contamination ranges
between $\sim$0 and $\sim0.4$ contaminants per arcmin$^{2}$ in the
magnitude range \jj=26.5-30. 
As already mentioned, the relative depth of the GNw seems to do the best job in
discriminating interlopers and dropouts, therefore the contamination is the smallest.

%%%%%%%%%%%%%%%%%%%%%
\begin{figure*}
\begin{minipage}{7in}
  \centering
%  \raisebox{-0.4\height}{\includegraphics[height=1.3in]{i_102.pdf}}
%  \raisebox{-0.3\height}{\includegraphics[height=1.in]{GSDZ-2130749007_sect33_f850lp_perpaper.pdf}}
%  \raisebox{-0.4\height}{\includegraphics[height=1.3in]{i_29.pdf}}
%  \raisebox{-0.3\height}{\includegraphics[height=1.in]{GSDZ-2388149536_sect23_f850lp_perpaper.pdf}}
  \raisebox{-0.4\height}{\includegraphics[height=1.3in]{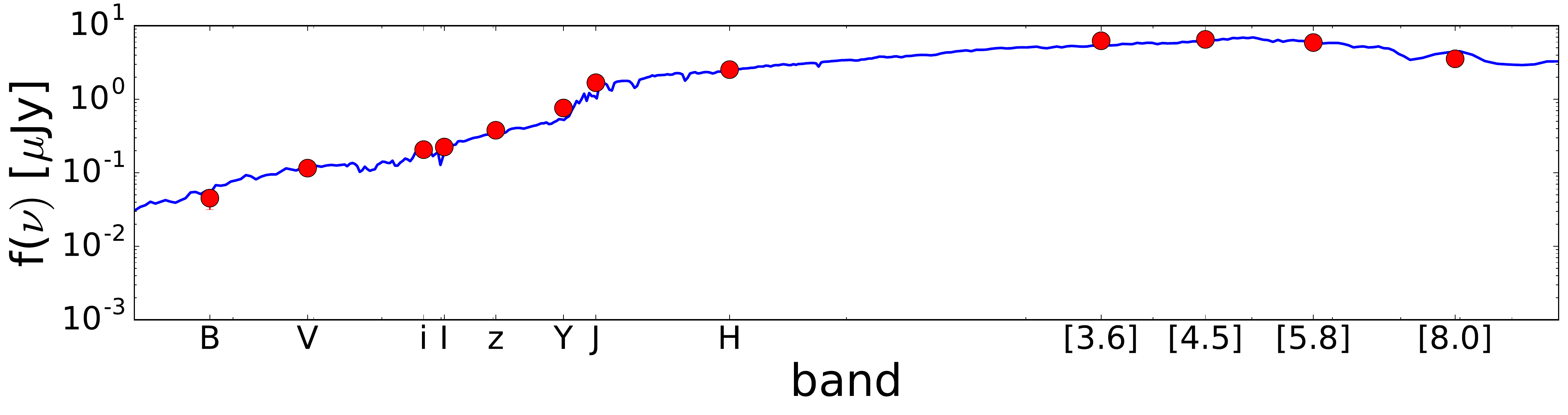}}
  \raisebox{-0.3\height}{\includegraphics[height=1.in]{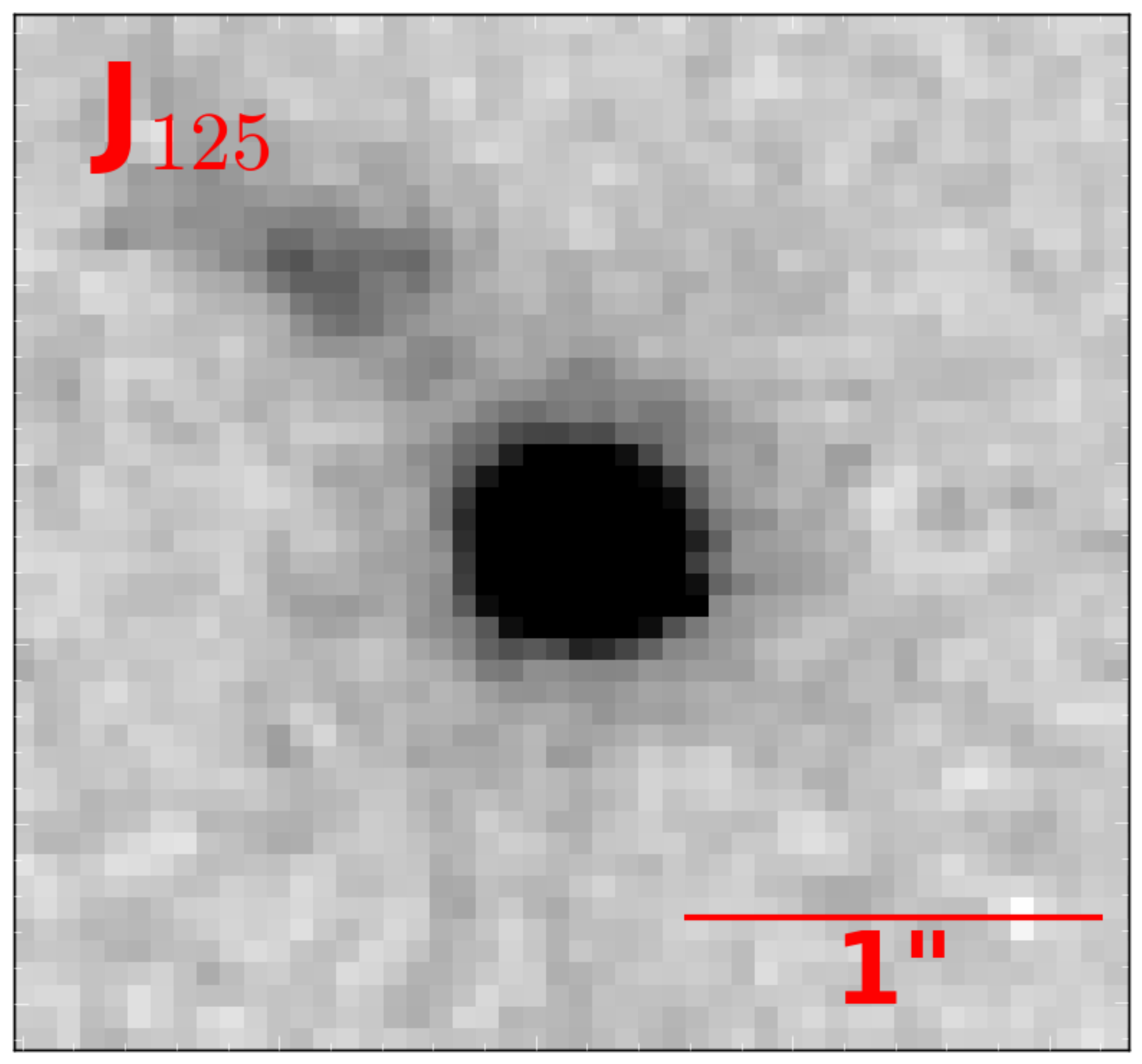}}
  \raisebox{-0.4\height}{\includegraphics[height=1.3in]{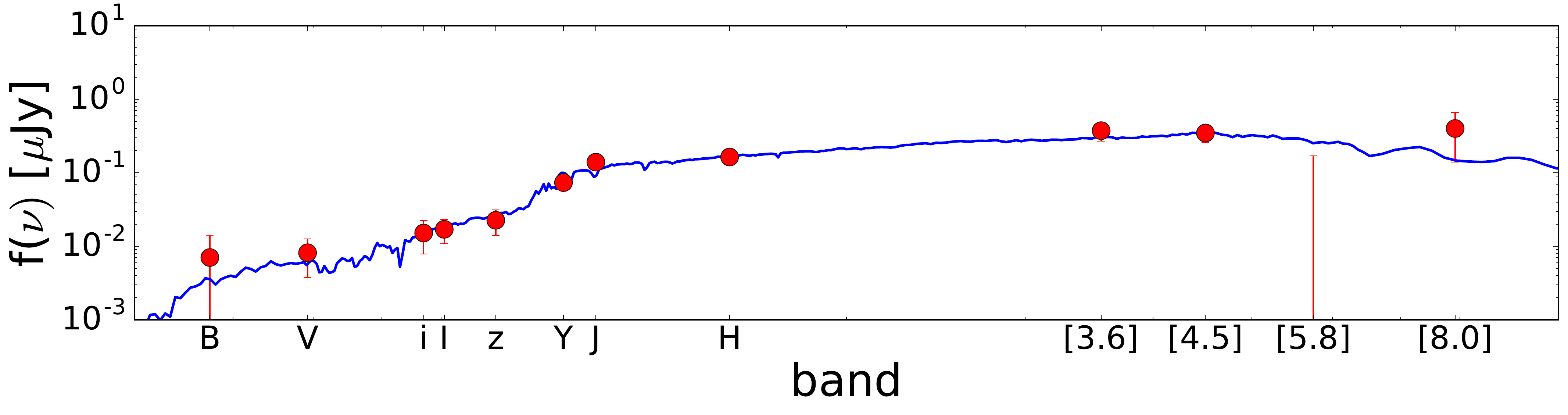}}
  \raisebox{-0.3\height}{\includegraphics[height=1.in]{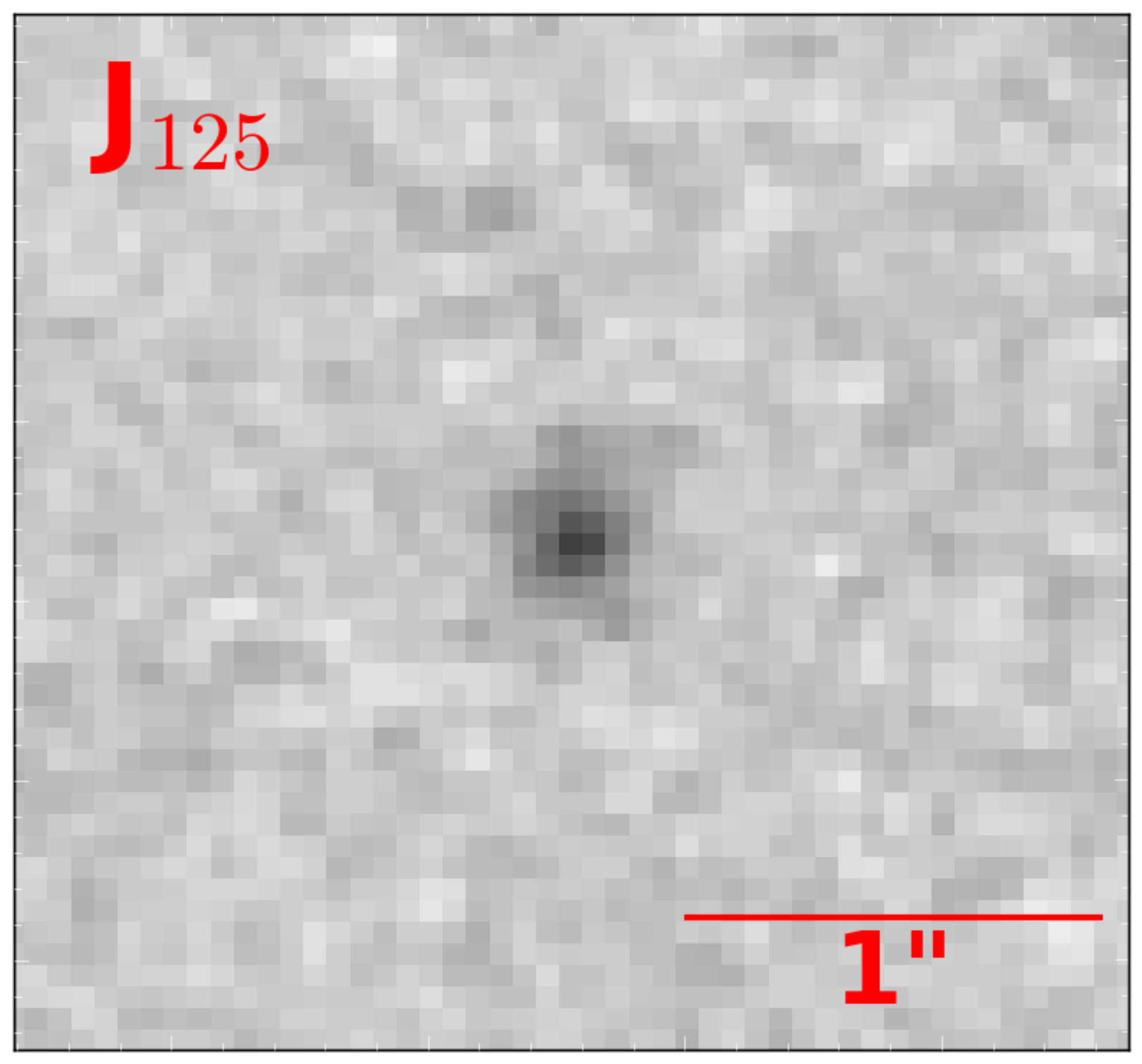}}
\end{minipage}
\caption{Examples of SEDs and images in the band of detection for  \yy-interlopers that  appear as dropouts at  $z\sim8$ (right) after the MC experiment. SED fitting has been obtained the photometry with FAST \citep{kriek2009}.} \label{Fig:images_interlopers}
\end{figure*}
%%%%%%%%%%%%%%%%%%%%%

Our conclusion on the presence of a significant level of contamination
near the detection limit of a survey because of significant
photometric scatter is indirectly supported by a cross-matching
analysis of the catalogs for \ii and \yy-dropouts in the XDF/GOODS-South
published by \citet{mclure2013} and \citet{bouwens2015}, which shows
that less than $50\%$ of the sources appear in both catalogs within
one magnitude of the survey detection limit, even though the derived
luminosity functions are similar (see \citealt{nugent2014}).

%%%%%%%%%%%%%%%%%%%%%

\subsection{Properties of the \yy-contaminants}

%%%%%%%%%%%%%
\begin{figure*}
\centering
\includegraphics[scale=0.38]{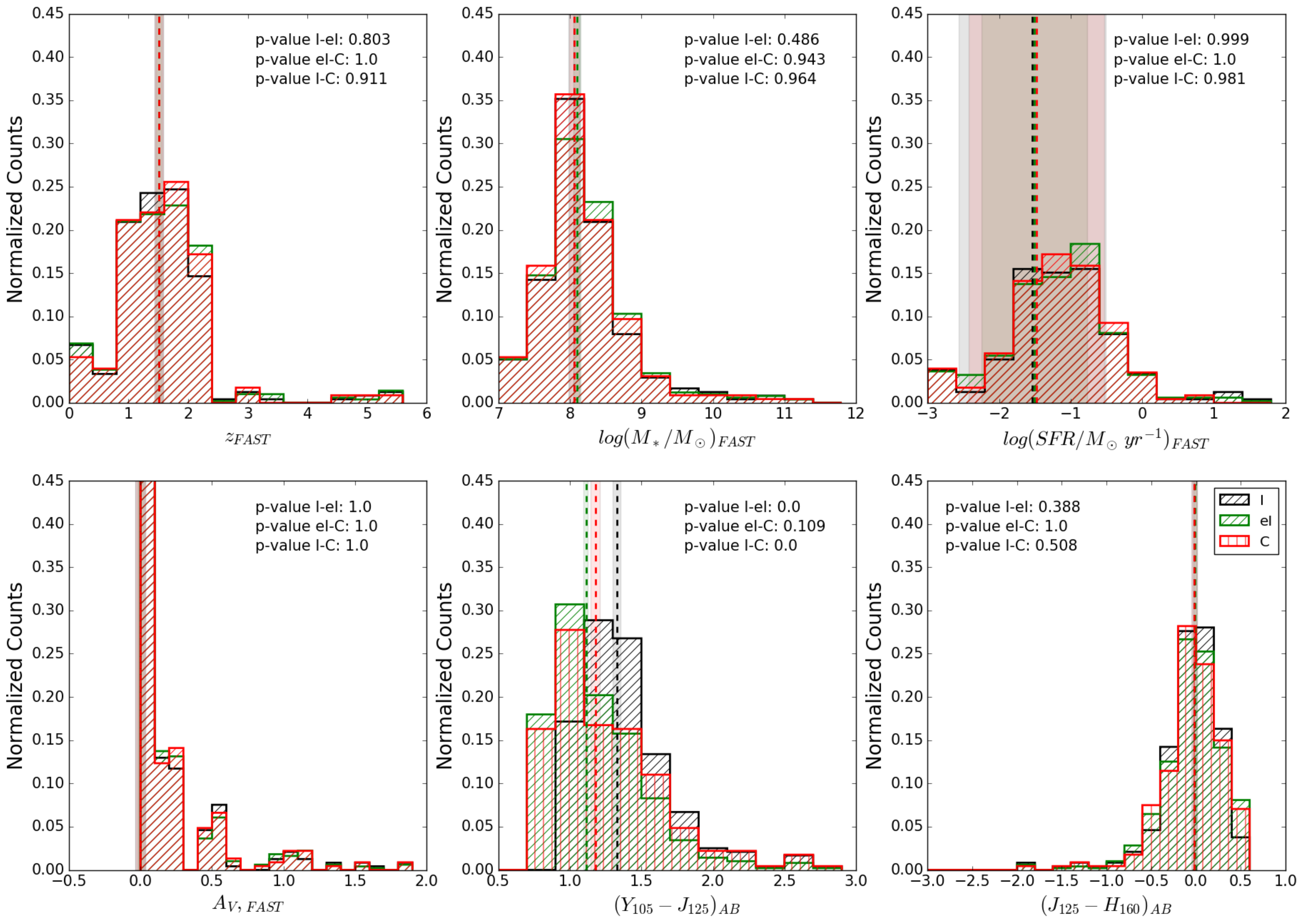}
\caption{Distribution of redshifts, stellar masses, star formation rates, dust extinctions and colors for Y-interlopers in the original sample (``I'', black lines), Y-interlopers in the enlarged  sample (``eI'', green lines), and for interlopers appearing as contaminants after the MC experiment (``C'', red lines). Stellar population properties have been estimated from the photometry with FAST \citep{kriek2009}. Values indicated are the probabilities that two distributions are drawn from the same parent distribution, according to the  Kolmogorov-Smirnov test. 
 %In the labels, numbers represent the number of objects in each sample. }
\label{Fig:inter_vs_cont}}
\end{figure*}
%%%%%%%%%%%%%

To investigate what are the properties of the objects that can migrate
from the interloper to the dropout sample when their photometry is
rescaled to fainter fluxes (and therefore lower S/N), we report in Figure
\ref{Fig:images_interlopers} some examples of interlopers in the
enlarged sample that after the MC dimming experiment appear as dropouts 
at $z\sim8$.  As it is clear from the SEDs, these objects are bright in the band of detection, but
their 4000 \AA{} is sufficiently deep that the faint flux at bluer
wavelengths is not detected after the typical dimming of $\sim 3-4$
mag that our MC experiment assigns to simulated objects near the XDF
detection limit. The figure also highlights the key
assumption (and potential limitation) of our approach, that is the use
of SEDs observed in brighter galaxies for
modeling the colors of fainter sources. 
 
To further characterize the interlopers and especially those that after the dimming appear as dropouts, 
we derive their stellar population properties by fitting the observed 
SEDs from the F435W to the F160W  or to the Spitzer-IRAC 8$\mu m$ photometry,\footnote{We resort to the IRAC photometry 
from CANDELS \citep{guo2013}, which we matched to our sources based on coordinates. IRAC photometry is available only for galaxies in GSd} depending on availability, using FAST \citep{kriek2009}.
We adopt \cite{bc2003} models assuming exponentially declining Star formation histories (SFHs) of the form $SFR \propto \exp{-t/\tau}$,
where $SFR$ is the star formation rate, $t$ is the time since the onset of star formation, and $\tau$ sets 
the timescale of the decline in the SFR, solar metallicity, a \cite{calzetti2000} dust law, and a \cite{chabrier2003} 
Inital Mass Function (IMF). We allow $\log (\tau /Gyr)$ to range between 7.0 and 10.0 Gyr, 
$\log (t /Gyr)$ between 7.0 and 10.1 Gyr, and $A_V$ between 0 and 4 mag.
When possible, we also use photometric redshifts from the 3D-\emph{HST} survey
\citep{skelton2014}, to further constrain the fits.  
 
 Overall, across the different field, 273 \yy-interlopers  appear as contaminants in at least one out of the 500  MC 
 realizations. We expect this sample to be representative of the entire contaminant population.
 
A summary of the typical properties of the interlopers and of those that might
contaminate the dropout samples at  $z\sim8$ is given in
Table \ref{tab:stellar_prop}. The distributions of some properties are also presented in Fig. \ref{Fig:inter_vs_cont}.  Interestingly,
both interlopers and contaminants have
intermediate ages, low level of ongoing star formation and only a
moderate dust content. Both medians value and  Kolmogorov-Smirnov tests support the similarity
of the distributions. As expected given the fact that our contaminants are drawn from 
the enlarged sample, which by construction includes objects up to 0.2 mag bluer  than interlopers, 
 interlopers have a  noticeably redder
 \yy-\jj color than contaminants. %However, this was  .  \bbv{talk}
These estimates are consistent with the typical values of dust content and ages obtained from the
contamination model based on source simulations from an extensive SED library  used in Sec. \ref{sec:contam}.
 This result suggests that it is
reasonable to expect that such properties can scale from the
intermediate mass objects used as templates to the lower mass and
fainter sources that would be contaminants in actual datasets. 

%%%%%%%%%%%%%%%%%%%%%%%
\begin{table}
\begin{center}
\caption{Stellar population properties of all \yy-interlopers and of those appearing as dropouts after the MC experiment. }
\label{tab:stellar_prop}
\begin{tabular}{llcc}
\hline
\hline
property 					& \yy-interlopers & \yy-contaminants \\
\hline
$z$		&1.51$\pm$0.07&1.51$\pm$0.07 \\
(\yy-\jj)$_{AB}$		&1.33$\pm$0.03&1.21$\pm$0.04 \\
(\jj-\hh)$_{AB}$		&-0.03$\pm$0.04&-0.02$\pm$0.04 \\
$\log (M_\ast/M_\sun)$		&8.07$\pm$0.08&8.06$\pm$0.06 \\
$\log (SFR/(M_\sun\, yr^{-1})$	&-1.0$\pm$1& -1.5$\pm$0.9\\
$\log (SSFR/yr^{-1})$		&-9.5$\pm$0.9& -9.5$\pm$0.8\\
$A_V$					&0.00$\pm$0.03&0.00$\pm$0.03\\
$\log (\tau/Gyr^{-1})$		&8.0$\pm$0.1& 8.0$\pm$0.1\\
$\log (t/Gyr^{-1})$		&8.60$\pm$0.04& 8.60$\pm$0.04\\
\hline
\end{tabular}
\tablecomments{Median values along with errors are listed.}
\end{center}
\end{table} 
%%%%%%%%%%%%%%%%%%%

%%%%%%%%%%%%%%%%
\section{Contamination estimates in the literature}\label{Sec:lit}
In the literature, there have been various studies that tried to give an estimate 
of the contamination in the dropouts sample, with the intent to correct the estimates
of the luminosity functions, but not to characterize the properties of the contaminants.
Each of these studies has used a different definition for the 
dropout/interloper sample and evaluated the contamination in a different way, so a direct comparison among the different findings is not always possible
and have to be considered carefully. 

Here, we present a summary of some important literature results and then we will redo our analysis using the same 
selection criteria adopted by \cite{bouwens2015}, with the aim of directly compare our and their findings. %However, we note that, given the different criteria
%adopted by the different studies, direct comparisons among the different contamination estimates are not always possible and have to be considered carefully. 

\cite{bouwens2015} have estimated  the impact of a scattering into a color selection windows owing to the impact of noise
 by repeatedly adding noise to the imaging data from the deepest fields, creating catalogs, and then attempting to reselect sources from these fields in exactly the same manner as the real observations. Sources that were found with the same selection criteria as the real searches in the degraded data but that show detections blueward of the break in the original observations were classified as contaminants.
They  estimated the likely contamination by using brighter, higher-S/N sources in the XDF to model contamination in fainter sources. 
They estimated a contamination rate of 2$\pm$1\%, 3$\pm$1\%, 6$\pm$2\%, 10$\pm$3\%, and 8$\pm$2\% at $z\sim4$, $z\sim5$, $z\sim6$, $z\sim7$, and $z\sim8$, respectively.

These  results are  in agreement to ours (see also Sec.\ref{Sec:dataset_Bow})
 and to those found in other recent selections of sources in the high-redshift universe \citep[e.g.,][]{giavalisco2004, bouwens2006, bouwens2007, bouwens2011_hudf09, wilkins2011, schenker2013}.

\cite{finkelstein2015} found instead larger values of contamination. They estimated the contamination by artificially dimming  lower redshift sources in their catalog, to see if the increased photometric scatter allows them to be selected as high-redshift candidates.
For sources with 25$<$\hh$<$27, they  estimated a contamination fraction of  4.5\%, 8.1\%, 11.4\%, 11.1\%, and 16.0\% at $z\sim4$, $z\sim5$, $z\sim6$, $z\sim7$, and $z\sim8$, respectively. For fainter sources with 26$<$\hh$<$29, the contamination fraction increased to  9.1\%, 11.6\%, 6.2\%, 14.7\%, and $<$4.9\% at $z\sim4$, $z\sim5$, $z\sim6$, $z\sim7$, and $z\sim8$, respectively.
These  fractions are in line with the estimates from the stacked probability distribution  curves \citep[e.g.][]{malhotra2005}.

\cite{casey2014}, by studying the space density of the potentially contaminating sources, found that dusty star-forming galaxies at $z < 5$ might  contaminate $z > 5$ galaxy samples at a rate of $<1\%$. Such fraction might increase when photometric scatter is applied to faint, red galaxies, making it easier for them to scatter into high-z samples  \citep{finkelstein2015}.

 To minimize the probability of contamination by low-redshift interlopers, the BoRG strategy was to impose a conservative non-detection threshold of 1.5$\sigma$ on the optical-band data \citep{bradley2012, trenti2011}.
In order to estimate the residual contamination,  from the \cite{bouwens2010} data reduction they first identified F098M dropouts with F125W$<$27 considering a version of the GOODS F606W image degraded to a 5$\sigma$ limit F606W = 27.2 to match the relative F125W versus F606W BoRG depth. They then checked for contaminants by rejecting F098M dropouts with $S/N > 2$ in either B, V, or i (at their full depth).  They  estimated approximately 30\% contamination, which is much higher to what we found, but in good agreement with the estimate based on the application of the color selection to  libraries of SED models \citep{oesch2007}.

Note that  the key difference between BoRG and other surveys is that BoRG only has one blue band, making the identification of contaminants more difficult. 

\subsection{The Bouwens et al. 2015 cuts}
\subsubsection{Sample selection}
\label{Sec:dataset_Bow}
We now repeat our analysis adopting the cuts proposed by \cite{bouwens2015}, 
in order to test how a different sample selection may alter our conclusions. 
For the sake of brevity, we report our analysis performed only on the CANDELS/GOODS South deep imaging
\citep{grogin2011}.  The parent catalog is the one presented in \S\ref{Sec:dataset}.
We apply the same cut in  $S/N(JH_{\rm{det}})$ and stellarity index described in \S\ref{Sec:dataset}. 

We then apply the
following color selection criteria for samples of LBGs in the redshift
range $z\sim5-8$, based on \citet{bouwens2015}.

For $z\sim5$ candidates %(hereafter \vv-dropouts)
\begin{equation}
\begin{split}
V_{606}-i_{775}& >1.2  \\
z_{850}-H_{160}& <1.3 \\
V_{606}-i_{775}& >0.8(z_{850}-H_{160})+1.2
\end{split}
\end{equation}

For $z\sim6$ candidates %(hereafter \ii-dropouts)
\begin{equation}
\begin{split}
i_{775}-z_{850} & > 1.0\\
Y_{105}-H_{160}& < 1.0\\
i_{775}-z_{850}& > 0.78(Y_{105}-H_{160}) + 1.0
\end{split}
\end{equation}

For $z\sim7$ candidates %(hereafter \zz-dropouts)
\begin{equation}
\begin{split}
z_{850} - Y_{105} & > 0.7\\
J_{125} - H_{160} & < 0.45\\
z_{850} - Y_{105} & > 0.8(J_{125} - H_{160}) + 0.7
\end{split}
\end{equation}

For $z\sim8$ candidates %(hereafter \yy-dropouts)
\begin{equation}\label{eq:cuts}
\begin{split}
Y_{105} - J_{125} & >0.45\\
J_{125} - H_{160} & <0.5\\
Y_{105} - J_{125} & >0.75(J_{125} - H_{160})+0.525.
\end{split}
\end{equation}

\noindent To distinguish between interlopers and dropouts, we use the following cuts in S/N.   
According to \citet{bouwens2015},
\vv-dropouts are selected as sources with S/N(\bb)$<$2, \ii-dropouts
with S/N(\bb)$<2$ and either (\vv - \zz)$>$ 2.7 or S/N(\vv)$<2$, \zz-
and \yy dropouts with S/N($x$)$<2$ and $\chi^2_{x} <3$, where
 $x$ is intended to be  \bb, \vv, and \ii  bands for \zz-dropouts and \bb, \vv,  \ii and \i814  bands for \yy-dropouts (Eq.\ref{eq:chi2}). %  for the \zz dropouts and  \bb, \vv,  \ii  and \i814 bands for \yy-dropouts.
In addition, \zz-dropouts are also  selected as sources with  either
(\i814-\jj$)>$1.0 or  S/N(\i814)$<$1.5.  

As before, if a dropout satisfies more than one dropout selection, we
assign it to the highest redshift sample. This additional cut removes
$\sim 30$ sources from the \zz- selection, while in the other cases at
most a few sources are removed.  In contrast, we do not apply this
restriction to interlopers, which thus may enter multiple selections.
However, only very few galaxies enter simultaneously more than one selection, 
therefore results are not driven by this subpopulation of duplicates.

\subsubsection{Results}

%%%%%%%%%%%
\begin{figure}
\centering
\includegraphics[scale=0.18]{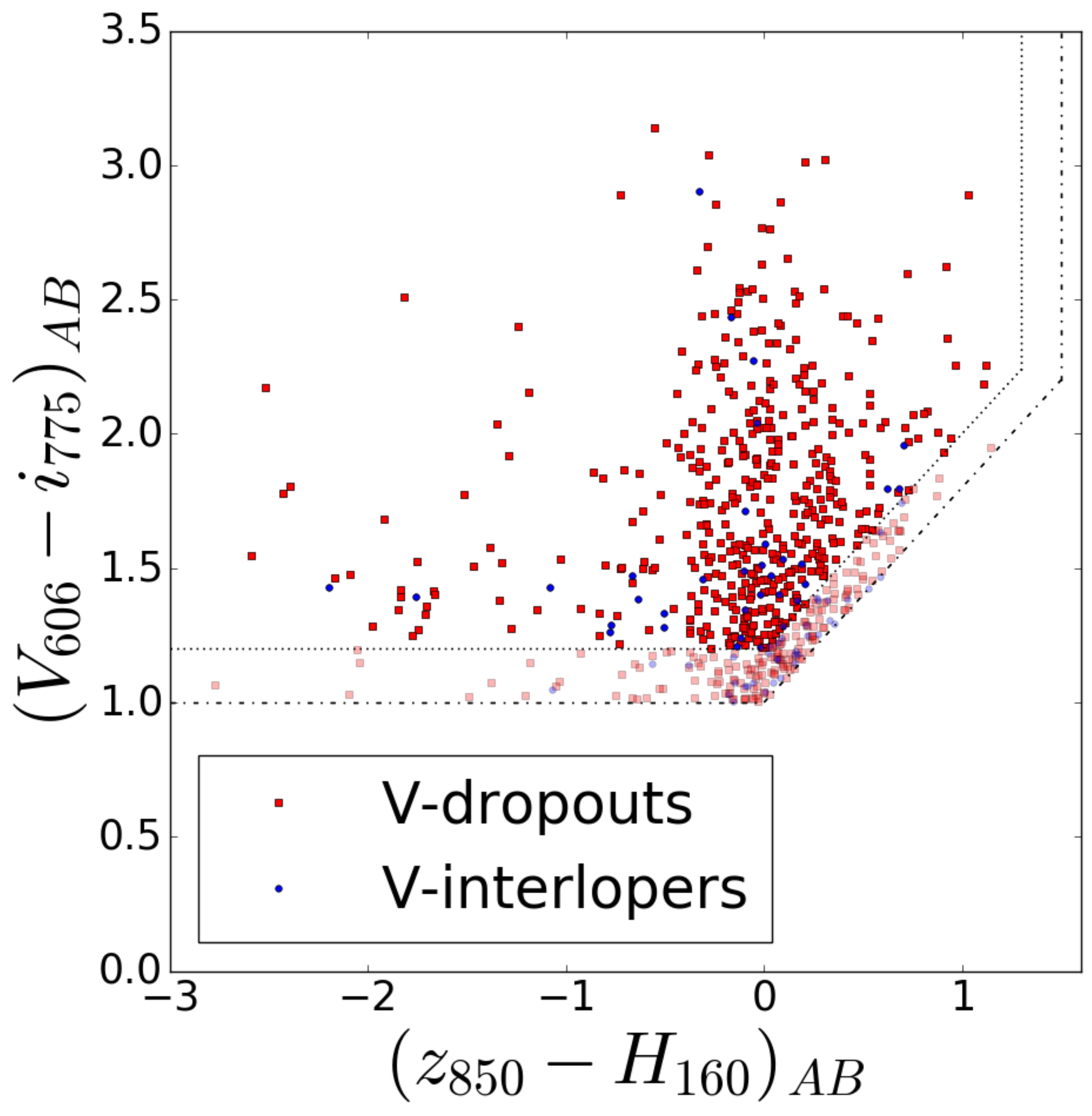}
\includegraphics[scale=0.18]{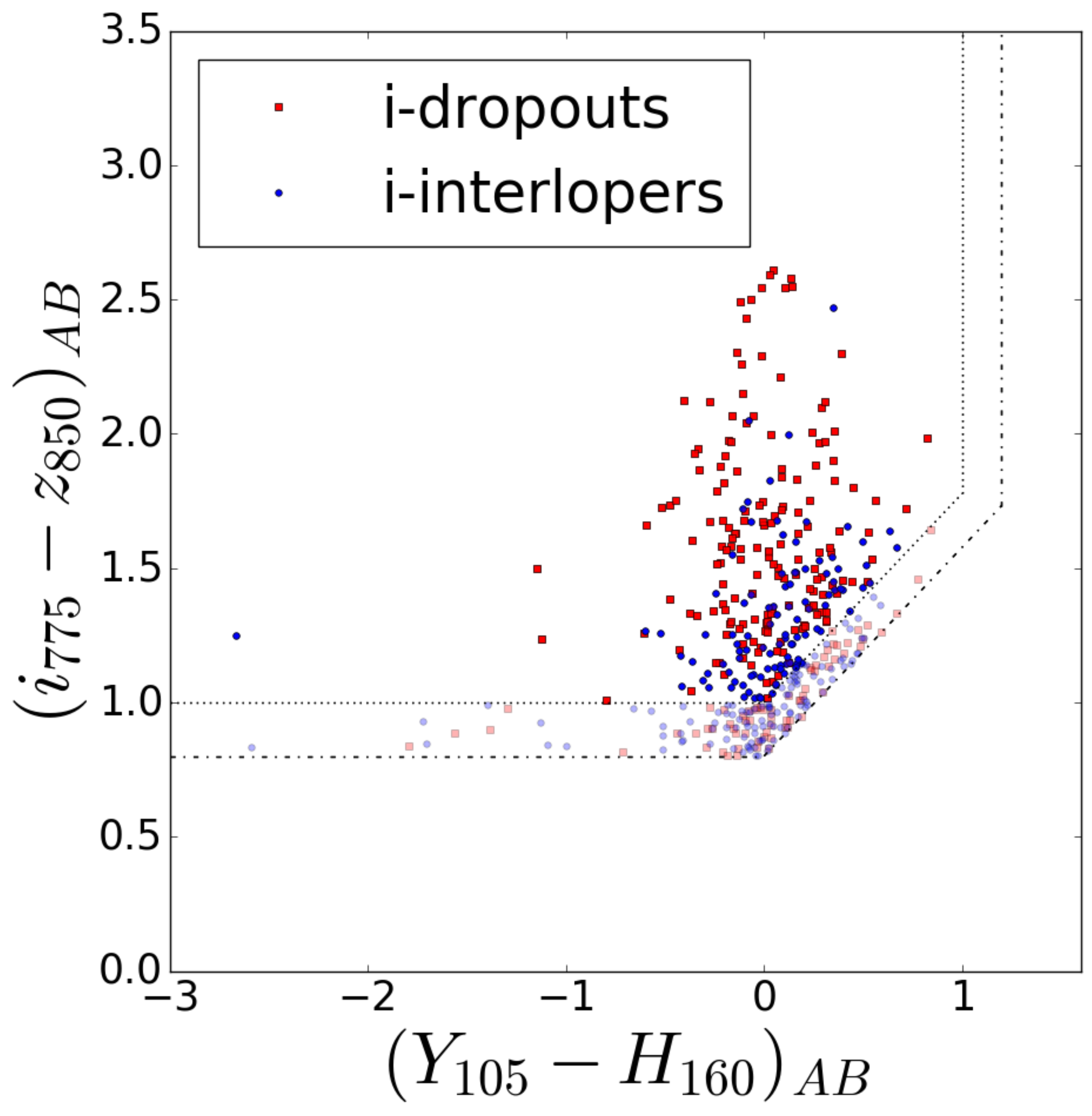}
\includegraphics[scale=0.18]{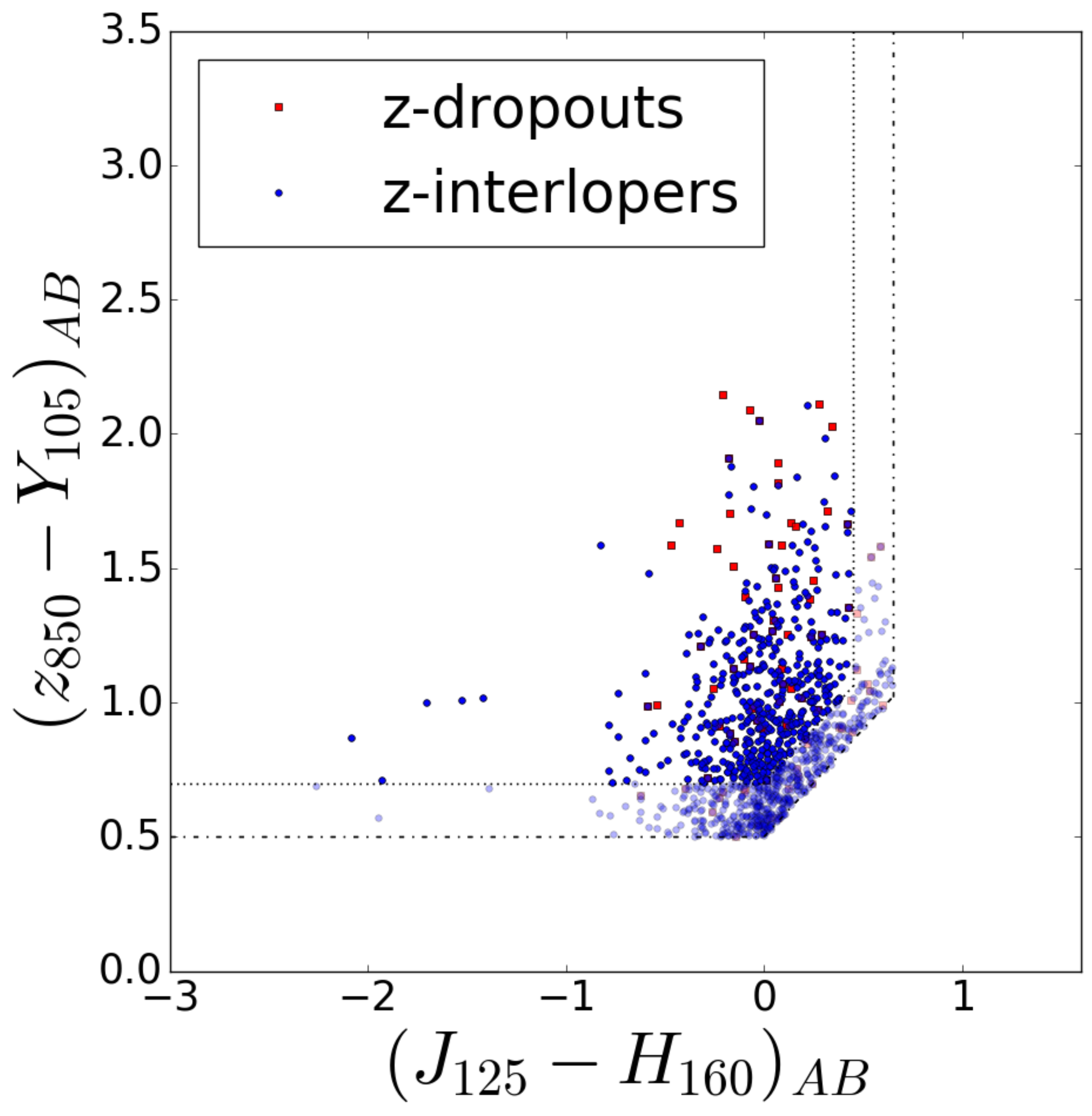}
\includegraphics[scale=0.18]{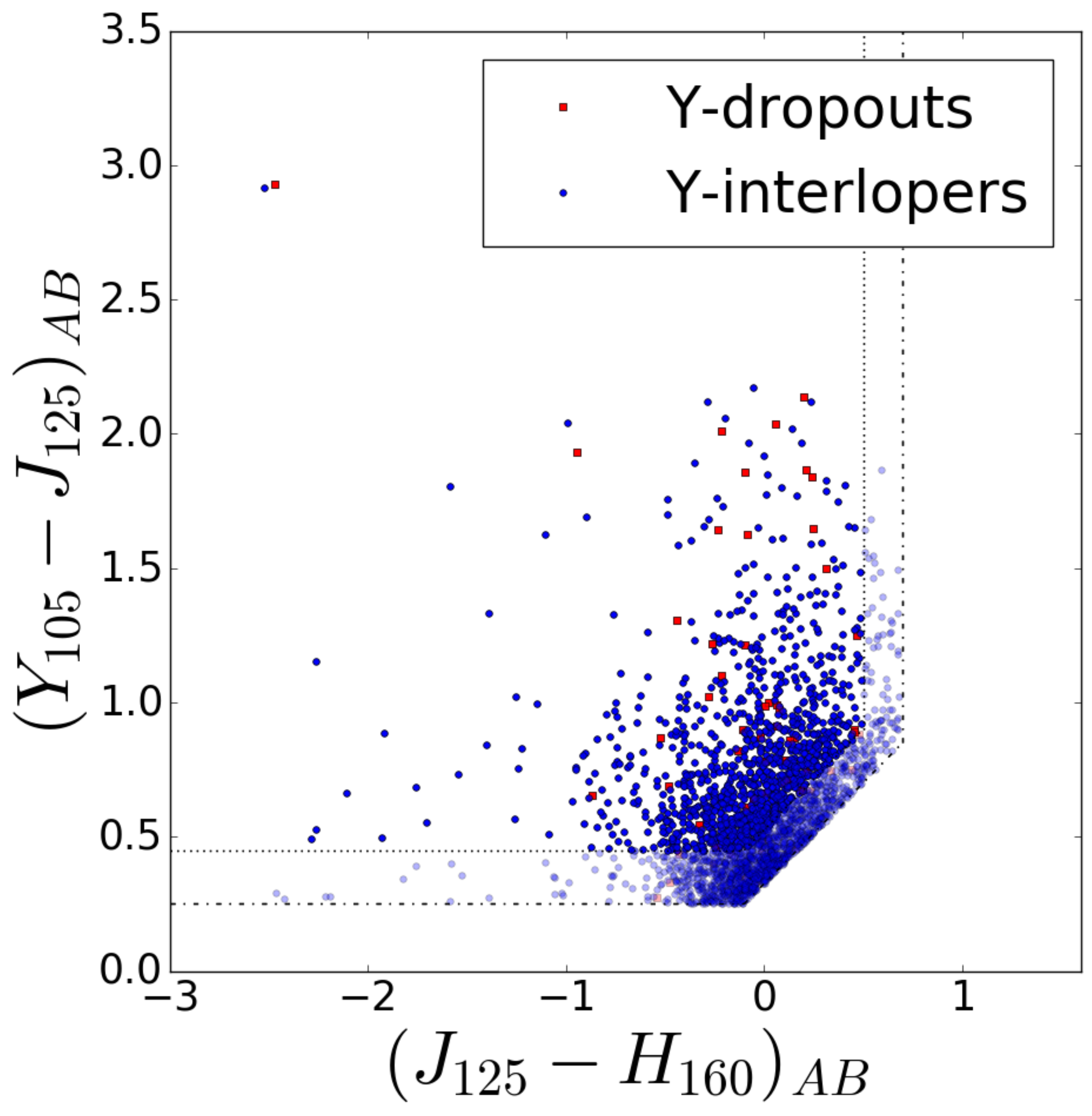}
\caption{Color-color selection box used to identify \vv- (upper left),
  \ii- (upper right), \zz- (bottom left) and \yy-dropouts (bottom
  right), following the color selection presented in \cite{bouwens2015}. % over the GOODS-SOUTH DEEP field. %(upper panel) and SHALLOW (bottom panel), respectively. For each sample,
  Red squares represent dropouts, i.e high-$z$ sources with no flux
  blueward of the Lyman-break; blue circles represent interlopers,
  i.e.  high-$z$ candidates showing a detection in the blue bands.
  Dashed lines represent the boundaries of the original sample
  selection, following \cite{bouwens2015}; dash-dotted lines represent
  the boundaries of the enlarged sample (see text for details). Darker
  symbols refer to the original selection, lighter ones to the
  enlarged
  selection.} %Stars are indicated with a starry symbol. \bbv{stars still need to be excluded from the sample!}}
\label{Fig:selection_box_B}
\end{figure}
%%%%%%%%%%

%%%%%%%%%%%%%%%%

%\subsection{Numbers and redshift distribution of dropouts and interlopers}

The color-color selection of dropouts and interlopers adopting the \cite{bouwens2015}
selection is shown in
Figure \ref{Fig:selection_box_B} for samples of \vv, \ii, \zz, and  \yy-dropout and interloper
sources. As done in \S 3.1, we  define an original and an enlarged selection, by simply
enlarging the color-color selection box by 0.2 mag, to check for both candidate high-$z$ LBGs
and interlopers that slightly fail to meet the usual selection criteria.

%%%%%%%%%%%%%%%%%%%%%%%
\begin{table}
\begin{center}
\caption{Statistics of dropouts and interlopers }
\label{tab:dropouts_percentages_B}
\begin{tabular}{l|cc|cc}
\hline
\hline
\multirow{2}{*}{population} & \multicolumn{2}{c|}{original sample}  & \multicolumn{2}{c}{enlarged sample}\\
     		& number & $\%$ & number & $\%$ \\    
\hline
\vv-dropouts & 446 & 93$\pm$2 & 601&90$\pm$2\\
\vv-interlopers & 33 &7$\pm$2 & 69 & 10$\pm$2\\
\hline
\ii-dropouts & 167 & 62$\pm$4 & 225&53$\pm$4\\
\ii-interlopers & 102 &48$\pm$4 & 215 & 47$\pm$4\\
\hline
\zz-dropouts & 53 & 11$\pm$2 & 74&7$\pm$1\\
\zz-interlopers & 443 &89$\pm$2 & 999& 93$\pm$1\\
\hline
\yy-dropouts & 45 & 4.1$\pm$0.9 & 61&2.5$\pm$0.5\\
\yy-interlopers & 1054 &95.9$\pm$0.9 & 2418 & 97.5$\pm$0.5\\
\hline
\end{tabular}
 \tablecomments{ Errors are defined as binomial errors \citep{gehrels86}.}
\end{center}
\end{table}
%%%%%%%%%%%%%%%%%%%%

%%%%%%%%%%%%%%%%%%%%%%%%%
\begin{figure*}
\centering
\includegraphics[scale=0.3]{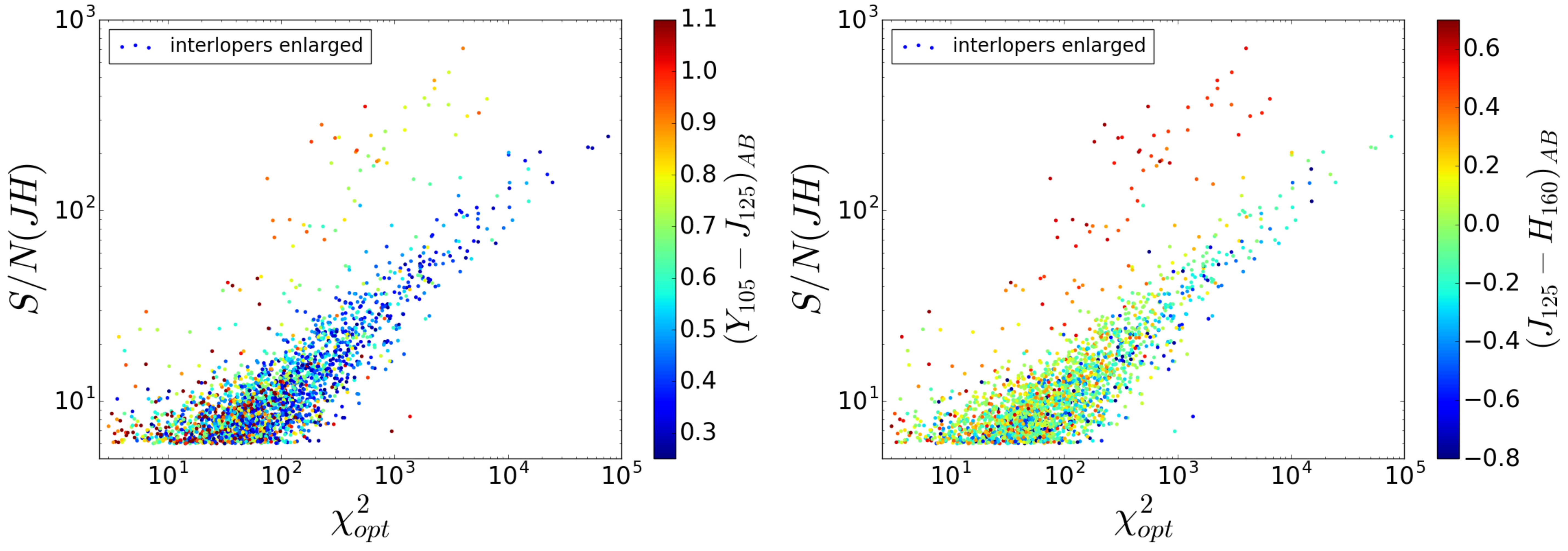}
\caption{Comparison between the ${\rm S/N(JH_{\rm{det}})}$ and the optical
  $\chi^2$ for interlopers in the enlarged sample at $z\sim8$. In the
  right panel, the \yy-\jj color is color-coded, while in the left
  panel the \jj-\hh color is color coded. }
 \label{Fig:SN2}
\end{figure*}
%%%%%%%%%%%%%%%%%%%%%%%%%%%%%%

Given the fact that the \cite{bouwens2015} criteria on the Lyman-Break color are less strict than those presented in \S1, many more galaxies enter both 
the dropout and the interloper samples, at any redshift.
Table~\ref{tab:dropouts_percentages_B} presents a summary of the incidence of each population. 
Comparing the fractions to those presented for the same field in Table~\ref{tab:dropouts_percentages}, 
we find that the fraction of \vv-dropouts is the same (the changes to the selection are really minor), while at higher
redshifts dropouts fraction are considerably smaller. This indicates that a more strict selection criteria does reduce the number
of interlopers, even though it simultaneously reduces the sample of dropouts. 
Therefore each selection should be a good compromise between 
purity and completeness.

Similarly to what we did in the previous section, we derive  the surface 
density distributions of dropouts and interlopers in our \cite{bouwens2015} like selection. 
Results are qualitatively similar to those found for a constant color cut, with 
the shape of the distribution of dropouts staying almost constant with increasing redshift, while that of interlopers considerably steepening.

Inspecting a ${\rm S/N(JH_{\rm{det}})}$ vs. $\chi^2_{opt}$ diagram for the  \cite{bouwens2015} 
sample selection, it emerges that the enlarged
sample appears to have objects distributed along two different
sequences. To further explore this population, in Fig. \ref{Fig:SN2}
we focus on interlopers only and add the information on their near-IR
colors. It appears evident that most of the objects in the second
sequence are characterized by intermediate colors in \yy-\jj and red
colors in \jj-\hh (0.5$<$\jj-\hh$<$0.7). This demonstrates the utility
of excluding candidates that are too red \citep[e.g.][]{giavalisco2004, bouwens2007, bouwens2015}.  Similar results also hold for samples from
selections at lower $z$.

\subsubsection{Contamination in dropout samples}\label{sec:contam}
Mimicking the analysis in the previous section, we investigate the level 
of contamination in the \cite{bouwens2015} dropout sample  induced by
interlopers that are mis-classified as dropouts in absence of  sufficiently 
deep data at bluer wavelengths. 

First we estimate  the impact of noise in the measurement of the optical
$\chi^2$ and photometric scatter in the color-color selection performing a
resampling MC simulation on the photometric
catalogs. As before, for each dropout selection, we uniformly
sample with repetition the luminosity in the detection band from the
catalog of enlarged interlopers, extracting a simulated catalog with
the same size as the original one. Next, we assign to each of these
objects the broadband colors of a random galaxy from the original
parent catalog (again using uniform sampling probability with
repetition), and we add zero-mean noise in the fluxes sampling from a
Gaussian distribution with width determined by the S/N ratio of the
simulated broadband fluxes. Finally, we perform the photometric
analysis of the catalog to quantify the number of interlopers in the enlarged sample that are
classified as dropouts. After repeating the procedure 500 times to
collect statistics, we find that on average:
%%%%
\begin{itemize}
\item The $z\sim 5$ selection has 7$\pm$1 interlopers entering the
  \vv-dropout sample as contaminant, for an estimated contamination rate
  $f_c\sim 7/446 \sim 1.5\%$;
\item The $z\sim 6$ selection has 4$\pm$1 interlopers entering the 
  \ii-dropout sample as contaminant, for an estimated contamination rate 
  $f_c\sim 4/167 \sim 2.5\%$;
\item The $z\sim 7$ selection has 5$\pm$1 interlopers entering the 
  \zz-dropout sample as contaminant, for an estimated contamination rate 
  $f_c\sim 5/53 \sim 9.4\%$;
\item The $z\sim 8$ selection has  7$\pm$1 interlopers entering the 
  \yy-dropout sample as contaminant, for an estimated contamination rate 
  $f_c\sim 7/45 \sim 15.3\%$.
\end{itemize}
%%%%%
These results are clearly illustrating that while the number of
mis-classified interlopers remains relatively constant across
different samples, as the redshift increases, the relative weight
compared to the number of dropouts grows significantly. These estimates
are systematically larger than those presented in the previous section, 
indicating how in the selection cuts proposed  \cite{bouwens2015} many more interlopers
might be incorrectly classified as dropouts. Nonetheless, 
as the sample presented in the previous section, these 
estimates are consistent with the predictions from the
contamination model based on source simulations from an extensive SED
library  \citep{oesch2007}. The model predicts a contamination of 3.1\% at
$z\sim5$, 0.5\% at $z \sim6$, 6.3\% at $z\sim7$ and 12.4\% at
$z\sim8$.  It forecasts a higher contamination at $z\sim5$
compared to $z\sim6$ that our MC experiment does not capture. This is
likely due to the fact that, being based on the observed data, the MC
at $z\sim5$ is able to identify contaminants only when the objects
show a detection in the single band (\bb) blueward of the break,
unlike the model based on a template library.

Therefore, it appears evident that the choice of the color cuts noticeably alters
the fraction of dropouts and interlopers and the estimates of contamination. 
The \cite{bouwens2015} selection criteria ensure a larger number of dropouts at all
redshifts, but unavoidably also a larger number of interlopers. 
As a consequence also the estimated contamination is considerably higher. 

\section{Summary and Conclusions}
\label{Sec:conclusions}

In this paper we investigated the contamination of photometrically
selected samples of high redshift galaxies. Our focus has been on the
widely adopted Lyman-break technique, using high quality multi-band
imaging from the Hubble CANDELS surveys (GOODS Deep South and 
GOODS Wide North), the XDF and the HUDF09-2. In our analysis we distinguished between
dropouts, that is sources that formally satisfy all the selection
criteria of LBGs, and interlopers, that is sources
with similar colors redwards of the spectral break, but showing a
detection at bluer wavelengths. Because of finite photometric
precision,  when no sufficiently deep data at bluer wavelengths
are available, a (small) fraction of interlopers can be mis-classified as
dropouts, and contaminate the selection. Hence we indicated these
objects as contaminants. 

The class of interlopers/contaminants that we studied is that of
intermediate redshift galaxies with a prominent 4000 \AA{}/Balmer break, which
are the most common among interlopers based on redshift estimates from
the 3D\emph{HST} survey (see Figure~\ref{Fig:zphot}). 

Our key results are the following: 

\begin{itemize}

\item Adopting a constant cut on the strength of the Lyman break across different redshifts, 
  the number counts of interlopers shows an increase in number with $z$ of at most a factor
  of 2. In contrast, in selections where the cut on  the strength of the Lyman break varies with 
  redshift \citep[e.g.][]{bouwens2015}, the number counts of interlopers increase significantly  from $z\sim5$ to $z\sim8$. 
  This suggests that cleaner samples of dropouts can
  be achieved by requesting a clear spectral break, which reduces the
  number of interlopers more significantly than the number of
  dropouts.

\item The surface density of interlopers in the sky remains
  approximately constant over the range of dropout selection windows
  considered in this study (that is dropouts from $z\sim 5$ to
  $z\sim 8$), for a given depth of the survey and for a uniform cut in
  the color containing the Lyman break. This is because the population
  of interlopers resides at lower redshift and its average redshift
  evolves more slowly (by a factor $\sim 0.3$) compared to that of the
  dropouts (see Equation~\ref{eq:zint} and
  Figure~\ref{Fig:zphot}). Thus, since the number of dropouts evolves
  rapidly with redshift, the ratio of interlopers to dropouts grows
  significantly with increasing redshift.

\item While the shape of the surface density distribution of  dropouts
  stays relatively constant with increasing redshift, that of interlopers possibly gets steeper. 
 Interlopers also tend to
  have a tail at the bright end.

\item Using a Monte Carlo resampling of the interloper population we
  estimate a contamination of the dropout samples in all the fields,
  ranging from $\sim 2\%$ at $z\sim 5$ to $\sim 6\%$ at $z\sim 8$ for
  the GSd field, with a clear trend of increasing contamination for
  higher redshift dropout samples. In the other fields, the
  contamination is similar, but systematically lower. The lowest level
  of contamination is found for GDw, indicating that having relatively
  deeper blue bands compared to red bands is the most effective toop
  to properly separate interlopers from dropouts.

\item Extrapolating with a power-law the interloper number counts
  distribution at the faint end to simulate ultradeep surveys, we
  derive that the contamination increases toward fainter magnitudes,
  and ranges from 0.1 to 0.4 contaminants per arcmin$^2$ at \jj=30,
  depending on the field considered. Generally we find that these
  contaminants are located near the detection limit of the survey.

\item By means of SED modeling, we characterized the stellar population
  properties of the interlopers that may contaminate the dropout
  sample, and found objects with intermediate ages ($\sim 1$ Gyr at
  $z\sim 1.5-2$), very-low levels of ongoing star formation, and
  relatively low dust content.

\end{itemize}

Our results and contamination estimates are limited by restrictions to
galaxy-like sources and to Gaussian noise. The former is not likely to
be an issue for space based observations with high angular resolution,
but it might affect ground-based surveys that do not have the ability
to discriminate between compact galaxies and stars. The assumption of
normally distributed errors is again likely to underestimate the
occurrence of rare, extreme events of photometric scatter, since data
are likely to have an excess of noise compared to a normal
distribution in their tails \citep{schmidt2014}. Thus, our
results are to be considered lower limits for the contamination of
dropout samples. Also, while we focused on Lyman-break selection, a
similar analysis would be expected to hold qualitatively if we had
considered photometric redshift estimates to construct the sample of
dropouts and interlopers, with the added complication of leaving more
degrees of freedom in defining the selection and the separation
between the two samples. 

Overall our key conclusion is that the dropout selection of high
redshift sources leads currently to samples with high purity, but the
purity degrades when the number of dropouts becomes much smaller than
the number of interlopers. We demonstrated this clearly for the
\yy-dropout sample from space observations over deep fields.  A
qualitatively similar conclusion on an increase of the contamination
fraction is expected to hold for ground-based surveys over large areas
as well, targeting the bright-end ($m\sim 24-26$) of the galaxy
luminosity function at high-$z$, since the relative number of
dropouts versus interlopers is significantly suppressed. However, in
this case the objects are so bright that targeted follow-ups such as
spectroscopic observations, should be able to discriminate between
high-$z$ sources and contaminants.

Finally, extrapolating our results to future surveys at $z>10$, we
highlight the need to consider carefully the contamination of the
dropout samples, since the number of objects expected at such early
times will be orders of magnitude smaller than the number of
interlopers with similar colors, and thus the contamination might
exceed $50\%$. Fortunately, in this respect, the capability of 
\emph{JWST} to observe efficiently at rest-frame
optical wavelengths for sources at $z>10$ will greatly help in
continuing to select samples of photometrically selected objects with
high purity, similar to the role played currently by Spitzer IRAC
imaging to validate samples of bright dropouts at $z\sim 8-10$
identified by \emph{HST} \citep{bouwens2015}.

\acknowledgments{  We thank  the anonymous referee for their insightful remarks 
that helped us to improve the paper. B.V. acknowledges the support from an Australian
  Research Council Discovery Early Career Researcher Award
  (PD0028506). This work was partially supported by grants ARC
  FT130101593, and \emph{HST}/GO 13767, 12905, and 12572.

Facilities: \facility{\emph{HST}(ACS), \emph{HST}(WFC3).}

\bibliographystyle{apj}

\bibliography{valebib}
\end{document}